\documentclass[twocolumn,superscriptaddress, amsmath,amssymb,aps,prb,floatfix]{revtex4-2}
\usepackage[usenames,dvipsnames]{color}

\usepackage{comment}
\usepackage{cancel}
\usepackage{ulem}
\usepackage{physics}
\usepackage{graphicx}
\usepackage{dcolumn}
\usepackage{bm}
\usepackage{gensymb}
\usepackage{tensor}
\usepackage{rotating}
\usepackage[percent]{overpic}
\usepackage{multibib}
\usepackage[breaklinks=true,colorlinks,citecolor=blue,linkcolor=blue,urlcolor=blue]{hyperref}
\begin{document}
\title{Theory of intrinsic acoustic plasmons in twisted bilayer graphene}
\date{\today}
\author{Lorenzo Cavicchi}
\affiliation{Scuola Normale Superiore, I-56126 Pisa,~Italy}
\author{Iacopo Torre}
\affiliation{Departament de F\'{i}sica, Universitat Polit\`{e}cnica de Catalunya, Campus Nord B4-B5, 08034 Barcelona,~Spain}
\affiliation{ICFO-Institut de Ci\`{e}ncies Fot\`{o}niques, The Barcelona Institute of Science and Technology, Av. Carl Friedrich Gauss 3, 08860 Castelldefels (Barcelona),~Spain}
\author{Pablo Jarillo-Herrero}
\affiliation{Department of Physics, Massachusetts Institute of Technology, Cambridge, Massachusetts,~USA}
\author{Frank H. L. Koppens}
\affiliation{ICFO-Institut de Ci\`{e}ncies Fot\`{o}niques, The Barcelona Institute of Science and Technology, Av. Carl Friedrich Gauss 3, 08860 Castelldefels (Barcelona),~Spain}
\affiliation{ICREA-Instituci\'{o} Catalana de Recerca i Estudis Avan\c{c}ats, Passeig de Llu\'{i}s Companys 23, 08010 Barcelona,~Spain}
\author{Marco Polini}
\affiliation{Dipartimento di Fisica dell'Universit\`a di Pisa, Largo Bruno Pontecorvo 3, I-56127 Pisa,~Italy}
\affiliation{ICFO-Institut de Ci\`{e}ncies Fot\`{o}niques, The Barcelona Institute of Science and Technology, Av. Carl Friedrich Gauss 3, 08860 Castelldefels (Barcelona),~Spain}
\begin{abstract}
We present a theoretical study of the intrinsic plasmonic properties of twisted bilayer graphene (TBG) as a function of the twist angle $\theta$ (and other microscopic parameters such as temperature and filling factor). Our calculations, which rely on the random phase approximation, take into account four crucially important effects, which are treated on equal footing: i) the layer-pseudospin degree of freedom, ii) spatial non-locality of the density-density response function, iii) crystalline local field effects, and iv) Hartree self-consistency. We show that the plasmonic spectrum of TBG displays a smooth transition from a strongly-coupled regime (at twist angles $\theta \lesssim 2^{\degree}$), where the low-energy spectrum is dominated by a weakly dispersive intra-band plasmon, to a weakly-coupled regime (for twist angles $\theta \gtrsim 2^{\degree}$) where an acoustic plasmon clearly emerges. This crossover offers the possibility of realizing tunable mid-infrared sub-wavelength cavities, whose vacuum fluctuations may be used to manipulate the ground state of strongly correlated electron systems.
\end{abstract}
\maketitle

\section{Introduction}\label{sect:intro}
Parallel two-dimensional electron systems (P2DESs) have been at the center of a great deal of attention since they were theoretically proposed in 1975 as ideal setups for the study of  superfluidity of spatially separated electrons and holes~\cite{lozovik_jept_1975}. They have been experimentally fabricated by using two main experimental platforms: i) one based on GaAs/AlGaAs heterostructures realized by molecular beam epitaxy~\cite{Dingle_APL_1978,Manfra_arxiv_2013,Chung_PRM_2020,Chung_NatMater_2021} and ii) one on atomically-thin 2D materials, such as graphene and transition-metal dichalcogenides (TMDs), produced by mechanical exfoliation~\cite{Geim_Grigorieva_Nature_2013}. These systems harbor a wide set of spectacular electrical phenomena, including Coulomb drag~\cite{pogrebinskii,price,Gramila_PRL_1991,gorbachev_natphys_2012,Rojo_JPCM_1999,Coulomb_drag_2016}, exciton superfluidity in strong~\cite{eisenstein_macdonald_nature_2004,Eisenstein_ARCMP_2014,su_natphys_2008,liu_natphys_2017,li_natphys_2017} and zero~\cite{burg_prl_2018} magnetic fields, and broken symmetry states~\cite{Feldman_NatPhys_2009,Weitz_Science_2010,Mayorov_Science_2011,Freitag_PRL_2012,Bao_PNAS_2012,Velasco_NatureNano_2012} driven by strong electron-electron interactions. 

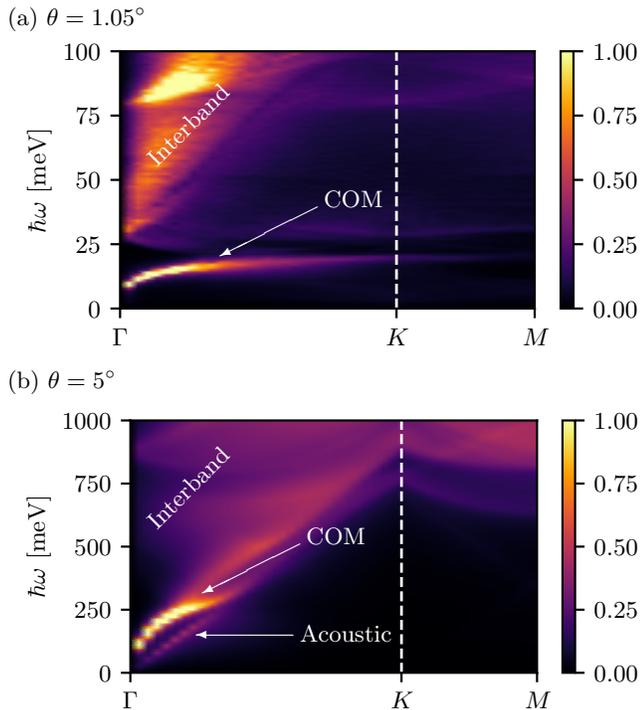
\begin{figure}[h!]
\centering
    \begin{overpic}[width=1\columnwidth]{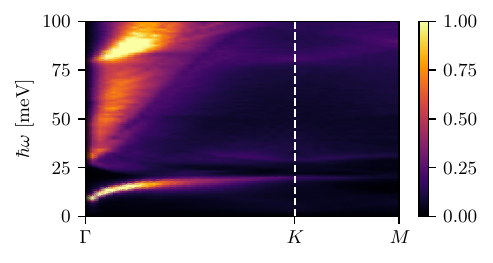}%
    \put(0,53){(a) $\theta = 1.05^\circ$}
    \begin{turn}{45}
    \put(30,14){\color{white} Interband}
    \end{turn}
    \put(36,25){\color{white}\vector(-2,-1){15}}
    \put(37,25){\color{white} COM}
    \end{overpic}\\
    \vspace{3mm}
    \begin{overpic}[width=1\columnwidth]{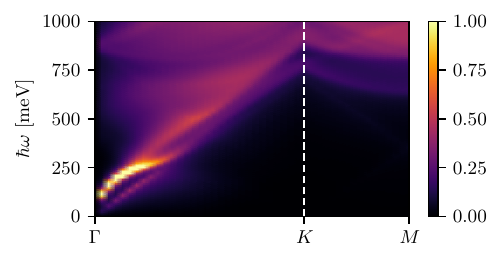}
    \put(0,53){(b) $\theta = 5^\circ$}
    \put(45,29){\color{white}\vector(-2,-1){15}}
    \put(46,29){\color{white} COM}
    \put(44,15){\color{white}\vector(-1,0){15}}
    \put(45,14){\color{white} Acoustic}
    \begin{turn}{45}
    \put(30,14){\color{white} Interband}
    \end{turn}
    \end{overpic}\\
    \vspace{0mm}
\caption{(Color online) The TBG energy loss function ${\cal L}({\bm q},\omega)$ as a function of ${\bm q}$ and $\omega$. The dependence on ${\bm q}$ is displayed along the high-symmetry path $\Gamma$-$K$-$M$ of the moir\'{e} BZ---see Fig.~\ref{fig:sketch+mBZ}(b). Results in this plot refer to filling factor $\nu=+1$ and temperature 
$T= 5~{\rm K}$. Panel (a) Results for $\theta = 1.05^\circ$ (chemical potential $\mu = 22$~meV). (b) Results for $\theta = 5^\circ$ (chemical potential $\mu = 256$~meV). In panel (b), an acoustic plasmon mode is clearly visible at low energies, just above the upper edge of the particle-hole continuum, i.e.~$\omega= v^{\star}_{\theta}q$, $v^{\star}_{\theta}$ being the reduced Fermi velocity---see Eq.~\eqref{eqn:normalized_fermi_velocity} below and also Section I of Ref.~\cite{SM}. High-energy interband plasmons have been discussed at length in Refs.~\cite{stauber_2016, novelli, hesp}.\label{fig:acoustic_appearence}}
\end{figure}

More recently, the many-body physics of P2DEs has been greatly enriched thanks to the discovery~\cite{Cao1,Cao2} of correlated insulators and superconductors in twisted bilayer graphene (TBG). TBG~\cite{lopes_prl_2007,shallcross_prl_2008,mele_prb_2010,li_naturephys_2010,shallcross_prb_2010,bistritzer_prb_2010,lopes_prb_2012,morell_prb_2010} is a P2DES comprising two graphene sheets on top of each other, separated by a vertical distance $d$ on the order on $\approx 0.3~{\rm nm}$, and rotated by a twist angle $\theta$. In this system, inter-layer tunneling changes significantly as a function of $\theta$, leading to a dramatic spectral reconstruction at a small, magic angle on the order of $\approx 1.1\degree$~\cite{bistritzer_pnas_2011}. At this angle, the (moir\'e superlattice) Brillouin zone is covered by a pair of very weakly dispersing (so-called) ``flat bands" centered on the charge neutrality point~\cite{morell_prb_2010,bistritzer_pnas_2011}. The reduction of kinetic energy due to band flattening strengthens the role of electron-electron interactions and is believed to be responsible for the exciting many-body physics that has been experimentally unveiled (for recent reviews see, for example, Refs.~\cite{Andrei2021,Kennes2021}).

P2DESs are also intriguing setups from the point of view of their plasmonic properties, which have been studied theoretically since the Eighties~\cite{das_sarma_1981,santoro_giuliani}. Indeed, a single 2DES displays a {\it plasmon} mode~\cite{GiulianiVignale}, which, in the long wavelength $q\to 0$ limit, can be interpreted as a center-of-mass (COM) oscillation dispersing as $\omega_{\rm COM}(q)\propto \sqrt{q}$, as a function of the in-plane wave vector $q$. This mode is extremely well understood and its small-$q$ behavior is highly constrained by 2D electrodynamics~\cite{GiulianiVignale}, posing practically no bounds on approximate theories for the 2D interacting many-particle problem. On the contrary, two P2DESs harbor an additional collective mode, which behaves very differently from the COM mode, depending on the amplitude of the inter-layer tunneling between the two layers where electrons roam. Let us consider a P2DES realized via a GaAs/AlGaAs double quantum well~\cite{Dingle_APL_1978,Manfra_arxiv_2013,Chung_PRM_2020,Chung_NatMater_2021}. If the barrier between the two quantum wells is sufficiently strong, the inter-layer tunneling amplitude---which in these systems is well described by a constant quantity typically dubbed $\Delta_{\rm SAS}$, physically representing the splitting between the symmetric and anti-symmetric states in the two adjacent wells---is negligible. In this weak inter-layer tunneling  (i.e.~$\Delta_{\rm SAS}\to 0$) limit, the additional collective mode is acoustic~\cite{das_sarma_1981,santoro_giuliani}, i.e.~$\omega(q)\propto q$ for $q\to 0$. Viceversa, in the limit of strong inter-layer tunneling, the additional collective mode is gapped~\cite{Dassarma_prl_1998}, $\omega(q)\propto \Delta_{\rm SAS}$ for $q\to 0$. The many-body theory of this mode, either for $\Delta_{\rm SAS}=0$~\cite{santoro_giuliani} or $\Delta_{\rm SAS}\neq 0$~\cite{Abedinpour_prl_2007}, is much more subtle than that needed to describe the COM plasmon in a single 2DES. Gapless, acoustic plasmons exist also in graphene double layers and topological insulator thin films~\cite{profumo}, provided that the two P2DESs there hosted are well isolated so that inter-layer tunneling can be neglected. 

This Article focuses on a simple question. How is TBG ``placed'' in this general context? This question is motivated by the qualitative difference between the two inter-layer tunneling Hamiltonians in the systems mentioned above, i.e.~TBG and GaAs double quantum wells. While in the latter a constant tunneling $\Delta_{\rm SAS}$ works very well, in the former inter-layer tunneling is highly modulated in space on the moir\'{e} superlattice length scale. Moreover, TBG too consists of two layers and in principle should support two collective modes at low energies. However, at small twist angles near the magic angle, only one low-energy COM plasmon mode $\omega_{\rm COM}(q)\propto \sqrt{q}$ is seen in state-of-the-art theoretical calculations of the plasmonic modes of TBG~\cite{stauber_2013,stauber_2016,levitov,novelli}. Where is the acoustic plasmon mode? 

The technical point is that in order to find an {\it intrinsic} acoustic plasmon in TBG~\cite{footnote_metal_gate}, one needs to deal with the layer-pseudospin degree of freedom. This needs to be included into the theoretical treatment of the plasmonic response of TBG, while at the same time taking into account three other important physical effects, namely spatial non-locality of the density-density response function beyond the Drude limit~\cite{santoro_giuliani,profumo,footnote_local_approximation}, Hartree self-consistency~\cite{guinea_2018,novelli} and crystalline local field effects~\cite{polini_2014,jung_2019}.

Accurate theoretical predictions for the plasmonic modes of TBG are important for a variety of fundamental and applied reasons. 
On the one hand, plasmons in TBG have been suggested as potential candidates for the microscopic explanation of superconductivity~\cite{Sharma_PRR_2020}. On the other hand, plasmon polaritons in TBG (and many other twisted 2D materials either with itinerant carriers or long-lived phonon modes) enrich the polariton panorama~\cite{Basov_Nanophotonics_2021}, providing us with a system with ultra-slow acoustic plasmons---see Sect.~\ref{sec:numerical}. 
Finally, since acoustic plasmons carry an electromagnetic field that is very well confined between the two layers~\cite{Lundeberg_Science_2017,Woessner_ACS_2017,AlcarazIranzo_Science_2018,Principi_PRB_2018}, they may have important applications in the field of quantum nanophotonics~\cite{Plantey_ACS_2021} and cavity QED of strongly correlated electron systems~\cite{GarciaVidal_Science_2021,Genet_PT_2021,Schlawin_APR_2022}. 
\begin{figure}[t]
\begin{overpic}[width=0.5\textwidth]{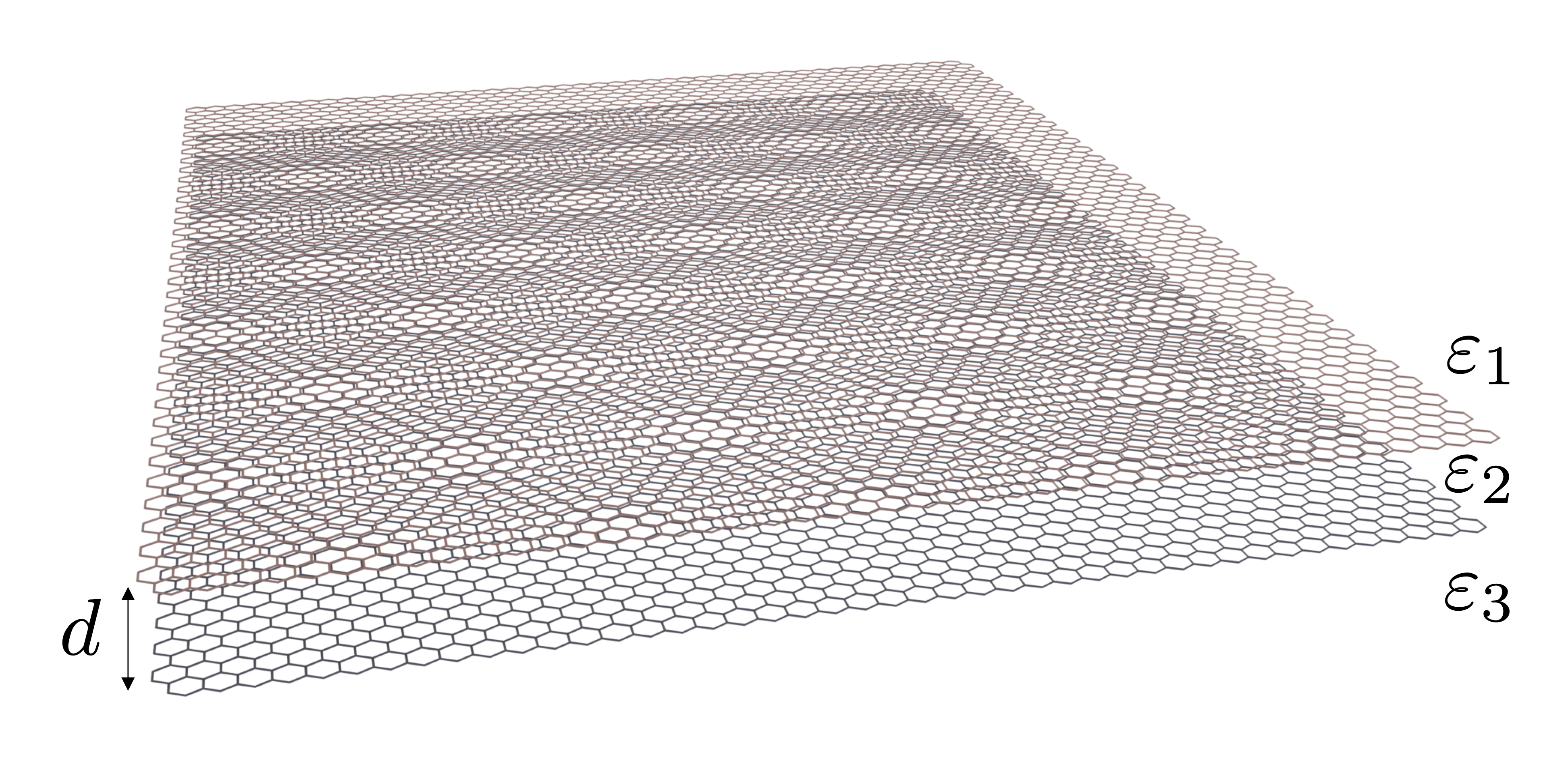}%
\put(0,43){(a)}
\end{overpic}
\begin{overpic}[width=0.5\textwidth]{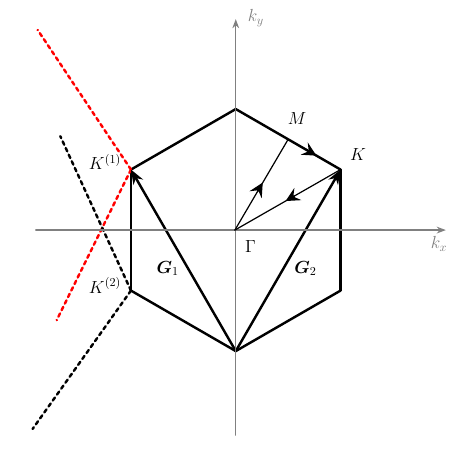}%
\put(0,80){(b)}
\end{overpic}
\caption{(Color online) (a) Sketch of the setup studied in this work.  TBG (spatial separation  between the two graphene layers denoted by $d$) is embedded in a dieletric environment described by three isotropic and homogeneous dielectrics with dielectric constants, $\varepsilon_1$ (top), $\varepsilon_2$ (middle), and $\varepsilon_3$ (bottom). (b) The first moir\'e BZ of TBG. The red (black) dashed lines are the edges of the BZ of the graphene layer ``1'' (``2''), $ K^{(1)} $ ($ K^{(2)} $) being the corresponding $ K $ point. The path $K$-$\Gamma$-$M$-$K$ is highlighted.}
\label{fig:sketch+mBZ}
\end{figure}

This Article is organized as following. In Sect.~\ref{sez:response} we introduce linear response theory for a P2DES consisting of two layers, formulating it for a system with in-plane Bloch translational invariance. In Sect.~\ref{sec:plasmonics} we summarize the theoretical approach we have used in this work, which we dub ``crystalline'' random phase approximation, introducing local field effects and the experimental observable we focus on, i.e.~the energy loss function. Section~\ref{sec:TBG_model} is devoted to a brief summary of the TBG continuum model Hamiltonian we rely on. Finally, in Sect.~\ref{sec:numerical} we present our main numerical results. Section~\ref{sect:summary} contains a brief summary and our main conclusions.
Sections~I-V of the Supplemental Material~\cite{SM} contain a wealth of additional numerical results. In particular, Sect.~IV deals with the role of an applied perpendicular electric field while Sect.~V discusses the impact of heterostrain.

\section{Linear response theory for two-layer P2DESs}\label{sez:response}
In this Section we summarize linear response theory (LRT)~\cite{GiulianiVignale} for a P2DES consisting of two layers. The formalism outlined here will be employed below in Sect.~\ref{sec:plasmonics} to evaluate the plasmonic spectrum of TBG. 

The ordinary density-density response function for a single 2DES~\cite{GiulianiVignale} can be easily extended to a P2DES consisting of two layers by using a $2\times2$ matrix formalism:
\begin{align}\label{eqn:bi_layer_response}
    \left(\begin{array}{c}
        \delta n^{(1)}(\bm q,\omega)   \\
        \delta n^{(2)}(\bm q,\omega)
    \end{array}\right) &= \int \frac{d^2\bm{q}^\prime}{(2\pi)^2}\left(\begin{array}{cc}
      \chi^{(1,1)}_{\hat{n}_{\bm{q}}\hat{n}_{-\bm{q}^\prime}}(\omega)   &  \chi^{(1,2)}_{\hat{n}_{\bm{q}}\hat{n}_{-\bm{q}^\prime}}(\omega)\\
       \chi^{(2,1)}_{\hat{n}_{\bm{q}}\hat{n}_{-\bm{q}^\prime}}(\omega)  & \chi^{(2,2)}_{\hat{n}_{\bm{q}}\hat{n}_{-\bm{q}^\prime}}(\omega)
    \end{array}\right)\nonumber\\
    &\times \left(\begin{array}{c}
        V_{\rm ext}^{(1)}(\bm{q}^\prime,\omega)   \\
        V_{\rm ext}^{(2)}(\bm{q}^\prime,\omega)  
    \end{array}\right)~.
\end{align}
Here, $\delta n^{(1)}(\bm q,\omega)$ and $\delta n^{(2)}(\bm q,\omega)$ are the Fourier components of the densities in the two layers, which are linked to the Fourier components of the two external scalar potentials  $V_{\rm ext}^{(1)}(\bm{q}^\prime,\omega)$ and $ V_{\rm ext}^{(2)}(\bm{q}^\prime,\omega)$ by a $2\times 2$ linear-response matrix. Its matrix elements are the quantities  $\chi^{(i,j)}_{\hat{n}_{\bm{q}}\hat{n}_{-\bm{q}^\prime}}(\omega)$, where $i,j=1,2$ are layer indices. For the sake of simplicity, we start by neglecting intra- and inter-layer electron-electron interactions. In this case, the off-diagonal elements $\chi^{(1,2)}_{\hat{n}_{\bm{q}}\hat{n}_{-\bm{q}^\prime}}(\omega)$ and $\chi^{(2,1)}_{\hat{n}_{\bm{q}}\hat{n}_{-\bm{q}^\prime}}(\omega)$ are non-zero only because of inter-layer tunneling, which couples layer $1$ with layer $2$ and viceversa. Electron-electron interactions will be included below in Sect.~\ref{sec:plasmonics}.

Good care needs to be exercised to correctly identify the layer-resolved density operators $\hat{n}_{\bm q}^{(i)}$ that lead to Eq.~(\ref{eqn:bi_layer_response}). The standard number density operator is defined by~\cite{GiulianiVignale} $\hat{n}(\bm{r}) = \sum_{k=1}^{N} \delta(\bm{r} - \hat{\bm r}_k)$, where the sum runs over the $k=1\dots N$ electrons. In a multi-layer structure, this operator is generalized to $\hat{n}^{(i)}(\bm{r}) = \hat{\Pi}^{(i)}{}^\dagger\hat{n}(\bm{r})\hat{\Pi}^{(i)}$. In the previous equation,  $i=1,2$ denotes the layer index and $\hat{\Pi}^{(i)}$ is the projector operator onto the $i$-th layer. In the case of two layers the total density operator is $\hat{n}(\bm{r}) =  \hat{\Pi}^{(1)}{}^\dagger\hat{n}(\bm{r})\hat{\Pi}^{(1)} +  \hat{\Pi}^{(2)}{}^\dagger\hat{n}(\bm{r})\hat{\Pi}^{(2)}$. An explicit construction of the projector operators is given below in Sec.~\ref{sec:TBG_model}. 

We now proceed to derive an expression for the quantity  $\chi^{(i,j)}_{\hat{n}_{\bm{q}}\hat{n}_{-\bm{q}^\prime}}(\omega)$, which applies to the case in which the P2DES is a crystal, i.e.~a Bloch translationally-invariant system. In this case, the single-particle eigenstates are of the Bloch type, i.e.~they are labeled
by a crystal momentum $\bm{k}$ belonging to the first Brillouin Zone (BZ) and a band index $\lambda$. A Bloch state $|\bm{k},\lambda\rangle$, with eigenvalue $\epsilon_{\bm{k},\lambda}$, is explicitly given by:
\begin{equation}\label{eq:single-particle-Bloch-state}
    \langle \bm{r} | \bm{k},\lambda \rangle = \frac{1}{\sqrt{S}}\sum_{\bm{G}} u_{\bm{G}}(\bm{k},\lambda) e^{i(\bm{k}+\bm{G})\cdot\bm{r}}~,
\end{equation}
where $S$ is the P2DES's area and $\bm{G}$ denotes the reciprocal lattice vectors of the crystal. Then, the  elements $\chi^{(i,j)}_{\hat{n}_{\bm{q}}\hat{n}_{-\bm{q}^\prime}}(\omega)$ of the non-interacting density-density response matrix can be expanded in a Bloch basis and the wave vectors $\bm{q}$ and $\bm{q}^\prime$ appearing in Eq.~\eqref{eqn:bi_layer_response} can differ at most by a reciprocal lattice vector (due to the periodicity of the lattice~\cite{GiulianiVignale, novelli}):
\begin{align}\label{eq:density-density_xstal_layer}
&\chi_{\hat{n}_{\bm{q}+\bm{G}}\hat{n}_{-\bm{q}-\bm{G}^\prime}}^{(i,j)}(\omega) =\nonumber\\
&= g_{\rm s} \int_{\rm BZ} \frac{d^2{\bm k}}{(2\pi)^2}
\sum_{\lambda, \lambda^\prime}\frac{f_{\bm{k},\lambda}-f_{\bm{k}+\bm{q}-\bm{Q},\lambda^\prime}}{\epsilon_{\bm{k},\lambda}-\epsilon_{\bm{k}+\bm{q}-\bm{Q},\lambda^\prime}+\hbar\omega+i\eta}\nonumber\\
&\times\langle \bm{k},\lambda|\hat{n}^{(i)}_{{\bm q} + \bm{G}}|\bm{k}+\bm{q}-\bm{Q},\lambda^\prime\rangle
\nonumber\\
&\times \langle \bm{k}+\bm{q}-\bm{Q},\lambda^\prime|\hat{n}^{(j)}_{-{\bm q} - \bm{G}^\prime}|\bm{k},\lambda \rangle~.
\end{align}
Here, $g_{\rm s}=2$ is a spin degeneracy factor,  $f_{\bm{k},\lambda}$ is the usual Fermi-Dirac distribution at chemical potential $\mu$ and temperature $T$,
\begin{equation}\label{eq:Fermi-Dirac}
f_{\bm{k},\lambda} = \frac{1}{\exp[(\epsilon_{{\bm k},\lambda}-\mu)/(k_{\rm B} T)]+1}~,
\end{equation}
and $\eta\to0^+$ is a positive infinitesimal. A folding vector $\bm{Q}$ belonging to the reciprocal lattice has been introduced in Eq.~(\ref{eq:density-density_xstal_layer}) to ensure that $\bm{k}+\bm{q}$ remains in the first BZ. 
\section{``Crystalline'' Random Phase Approximation}
\label{sec:plasmonics}

Plasmons are self-sustained density oscillations that emerge due to electron-electron interactions~\cite{GiulianiVignale}. These need to be treated at some level of approximation. Here, we employ the time-dependent Hartree approximation~\cite{GiulianiVignale}, also known as random phase approximation (RPA), and focus our attention on the electron energy loss function ${\cal L}(\bm{q},\omega)$. This quantity represents the probability of exciting the electronic system through the application of a scalar perturbation with wave vector $\bm{q}$ and energy $\hbar \omega$. ${\cal L}(\bm{q},\omega)$ contains valuable information about self-sustained charge oscillations, which appear as sharp peaks, as well as incoherent electron-hole pairs, which induce a broadening of the peaks or, more in general, produce a broadly distributed spectral weight in the ${\bm q}$-$\omega$ plane. The energy loss function can be in principle measured via electron energy loss spectroscopy~\cite{egerton} and scattering-type near-field optical spectroscopy (see, for example, Refs.~\cite{Basov_Nanophotonics_2021,Lundeberg_Science_2017,Woessner_ACS_2017,AlcarazIranzo_Science_2018} and references therein). 

As stated in Sect.~\ref{sect:intro}, the loss function will be calculated by including {\it local field effects} (LFEs)~\cite{adler, wiser,polini_2014, jung_2019, lewandowski}, naturally arising out of the underlying crystalline nature of the system under study. This is very naturally accomplished by retaining the dependence of the quantity $\chi_{\hat{n}_{\bm{q}+\bm{G}}\hat{n}_{-\bm{q}-\bm{G}^\prime}}^{(i,j)}(\omega)$  in Eq.~(\ref{eq:density-density_xstal_layer}) on the reciprocal lattice vectors $\bm{G}$, $\bm{G}^\prime$.

Finally, many-body effects, in general, and plasmons, in particular, are sensitive to the dielectric environment surrounding the P2DES under investigation. In this Article, we assume that TBG is embedded between two homogeneous and isotropic dielectric media described by the dielectric constants $\varepsilon_1$ (top) and $\varepsilon_3$ (bottom)---see Fig.~\ref{fig:sketch+mBZ}(a). The space between the layers is filled by a third homogeneous and isotropic dielectric characterized by a dielectric constant $\varepsilon_2$. In a typical experimental setup, the space between the layers is just a vacuum gap ($\epsilon_2=1$) and TBG is encapsulated between two slabs of hexagonal Boron Nitride (hBN), which is a homogeneous and {\it anisotropic} dielectric (therefore beyond the isotropic model introduced above). Such hBN slabs host hyperbolic phonon polariton modes~\cite{Basov_Nanophotonics_2021}, which strongly couple to plasmons~\cite{tomadin_PRL_2015}. We have therefore deliberately decided to neglect such plasmon-phonon polariton coupling in order to access, once again, the {\it intrinsic} plasmon modes of TBG. Including hBN polaritons into the theory is straightforward and can be accomplished by following for example the theory of Ref.~\cite{tomadin_PRL_2015}.

The loss function can be calculated from the following expression:
\begin{equation}\label{eqn:loss_function_xstal_layer}
    {\cal L}(\bm{q},\omega) = -\Im\left\{\mbox{Tr}_{\rm L}\left[\varepsilon(\bm{q},\omega)^{-1}\right]_{\bm{G}=\bm{0},\bm{G}^\prime=\bm{0}}\right\}~,
\end{equation}
where $\varepsilon(\bm{q},\omega)$ is the dynamical dielectric function, which, in the present case, is a matrix with respect to layer indices and reciprocal lattice vectors. The trace $\mbox{Tr}_{\rm L}$ in Eq.~\eqref{eqn:loss_function_xstal_layer} is intended to be over the layer-pseudospin degrees of freedom. We emphasize that, in order to evaluate the loss function via Eq.~\eqref{eqn:loss_function_xstal_layer}, the matrix $\varepsilon(\bm{q},\omega)$ needs to be inverted {\it before} a) the trace over the layer degrees of freedom is taken and b) the ${\bm{G}=\bm{0},\bm{G}^\prime=\bm{0}}$ element is selected.

Returning on the importance of LFEs, we remind the reader that the $\bm{G}=\bm{0}$, $\bm{G}^\prime=\bm{0}$ element of the inverse of the dynamical dielectric matrix $\varepsilon(\bm{q},\omega)$ produces the so-called ``macroscopic'' dielectric function~\cite{adler,wiser} $\varepsilon_{\rm M}(\bm{q},\omega)$, which is defined through the following equation:
\begin{equation}\label{eqn:macroscopic_dielectric}
    \varepsilon^{-1}_{\rm M}(\bm{q},\omega) \equiv  \left[\varepsilon^{-1}(\bm{q}, \omega)\right]_{\bm{G}=\bm{0},\bm{G}^\prime=\bm{0}}~.
\end{equation}
Inverting $\varepsilon(\bm{q}, \omega)$ {\it first}, and {\it then} selecting the $\bm{G}=\bm{0}$, $\bm{G}^\prime=\bm{0}$ element, brings to the macroscopic dielectric function contributions from non-zero reciprocal lattice vectors, i.e.~$\bm{G}\neq\bm{0}$, $\bm{G}^\prime\neq\bm{0}$. In solids, such LFEs are not negligible. As a result, the macroscopic field, which is the average of the microscopic field over a region larger than the lattice constant (but smaller than the wavelength) is not equivalent to the effective or local field that polarizes the charge in the crystal~\cite{adler, wiser}. This phenomenon is expected to be more relevant in systems with significant charge inhomogeneities, like moir\'e materials and TMDs~\cite{lewandowski, cudazzo, rubio}. In particular, modifications to the plasmon dispersion relation induced by LFEs tend to be important near BZ edges~\cite{lewandowski}. Importantly, the authors of Ref.~\cite{lewandowski} have recently shown that the inclusion of LFEs on the plasmon dispersion relation is crucial to probe correlated states in twisted hetero-bilayers of TMDs. More precisely, they argue that a loss function different from the one introduced in Eq.~(\ref{eqn:loss_function_xstal_layer}) and calculated by {\it tracing over the reciprocal lattice vectors} gives profound information about the many-body properties of the moiré material under investigation. While this is certainly true, standard plasmonic probes~\cite{Basov_Nanophotonics_2021,Lundeberg_Science_2017,Woessner_ACS_2017,AlcarazIranzo_Science_2018} usually access the response of the system to long-wavelength perturbations. Experimentally, therefore, the loss function defined in Eq.~\eqref{eqn:loss_function_xstal_layer} seems the more appropriate one to interpret plasmonic experiments, as briefly pointed out by the authors of Ref.~\cite{lewandowski} too. 

We now comment on the role of the layer degrees of freedom. At a first superficial glance, one may be puzzled by the definition of the loss function we gave above in Eq.~(\ref{eqn:loss_function_xstal_layer}) and, in particular, by its ability to display peaks at the collective modes of the layered structure. Indeed, in a layered structure, plasmon modes are calculated by looking at the zeroes of the {\it determinant} of the layer-resolved dielectric tensor~\cite{das_sarma_1981, santoro_giuliani}. How can we reconcile these two seemingly different approaches to the collective modes of layered materials? 
The answer is that the trace of the inverse dielectric tensor with respect to the layer degrees of freedom is proportional to the reciprocal of the determinant over the same degrees of freedom, i.e.~$\mbox{Tr}_{\rm L}\left[\varepsilon(\bm{q},\omega)^{-1}\right]_{\bm{G}=\bm{0},\bm{G}^\prime=\bm{0}} \propto 1/{\rm det}_{\rm L}\left[\varepsilon(\bm{q},\omega)\right]_{\bm{G}=\bm{0},\bm{G}^\prime=\bm{0}}$. We therefore see that there is no contradiction between the usual approach~\cite{das_sarma_1981, santoro_giuliani} and our loss-function based approach.

\subsection{Approximate dynamical dielectric matrix}

While the definition in Eq.~(\ref{eqn:loss_function_xstal_layer}) is totally general, we now need to introduce a necessarily approximate model for the dynamical dielectric matrix $\varepsilon(\bm{q}, \omega)$, which includes electron-electron interactions. 

In the RPA~\cite{GiulianiVignale}, we have
\begin{align}
\label{eqn:dielectric_tensorLFE_layer}
\left[\varepsilon(\bm{q}, \omega)\right]_{\bm{G},\bm{G}^\prime}^{(i,j)} &= \delta^{(i,j)}\delta_{\bm{G},\bm{G}^\prime} \nonumber\\
&- e^2 \sum_{\ell}L_{\bm{G}}^{(i,\ell)}(\bm{q})\chi_{\hat{n}_{\bm{q}+\bm{G}}\hat{n}_{-\bm{q}-\bm{G}^\prime}}^{(\ell,j)}(\omega)~,
\end{align}
where $L_{\bm{G}}^{(i,j)}(\bm{q}) =L^{(i,j)}(\bm{q}+{\bm G}) $ is the Coulomb propagator relating the charge density fluctuations $\delta n^{(j)}_{\bm{q}+\bm{G}}(\omega)$ to the self-induced electrical potential, i.e.~$W^{(i)}_{\bm{G}}(\bm{q},\omega) = e^2 L_{\bm{G}}^{(i,j)}(\bm{q}) \delta n^{(j)}_{\bm{q}+\bm{G}}(\omega)$. 

The quantities $L^{(i,j)}(q)$ are given by~\cite{profumo}:
\begin{equation}\label{eqn:L11}
L^{(1,1)}(q) = \frac{4\pi}{qD(q)}[(\varepsilon_2 + \varepsilon_3)e^{qd} + (\varepsilon_2 - \varepsilon_3)e^{-qd}] ~,
\end{equation}
and
\begin{equation}
L^{(1,2)}(q) = L^{(2,1)}(q) = \frac{8\pi}{qD(q)}\varepsilon_2~,
\end{equation}
where
\begin{equation}
D(q) =  (\varepsilon_1 + \varepsilon_2)(\varepsilon_2 + \varepsilon_3)e^{qd} +  (\varepsilon_1 - \varepsilon_2)(\varepsilon_2 - \varepsilon_3)e^{-qd}~.
\end{equation}
The expression for the $L^{(2,2)}(q)$ component is obtained from Eq.~\eqref{eqn:L11} by interchanging $\varepsilon_3$ with $\varepsilon_1$. In the presence of hBN dielectrics, the Coulomb propagator acquires a frequency dependence~\cite{tomadin_PRL_2015}, $L_{\bm{G}}^{(i,j)}(\bm{q}, \omega)$, due to the strong dependence of the hBN dielectric permittivity tensor on frequency in the mid-infrared spectral range.

It is now time to pause for a moment and discuss about the statements we have made about the non-local nature of the calculations reported in this Article. In the so called ``local approximation'' for calculating the plasmon dispersion relation in a single 2DES, the density-density response function in equation~\eqref{eqn:dielectric_tensorLFE_layer} is approximated with its value in the so-called ``dynamical limit''~\cite{GiulianiVignale}, i.e.~in the limit $q\to 0$ and $\omega\gg v_{\rm F}^* q$, where $v_{\rm F}^* q$ represents the upper edge of the electron-hole continuum. This approximation is extremely well suited to calculate the leading order term of the dispersion relation $\omega_{\rm COM}(q)$ of the COM mode in the long-wavelength $q\to 0$ limit. However, it is very well known~\cite{santoro_giuliani,profumo} that such local approximation fails in predicting the correct acoustic plasmon dispersion, even in the long wavelength $q\to 0$ limit. This is why, in this Article, we have decided to retain the full dependence of $\chi_{\hat{n}_{\bm{q}+\bm{G}}\hat{n}_{-\bm{q}-\bm{G}^\prime}}^{(i,j)}(\omega)$ in Eq.~(\ref{eq:density-density_xstal_layer}) on the wave vector ${\bm q}$, without making the local approximation (i.e.~without taking the dynamical limit).

\section{TBG model Hamiltonian and Hartree self-consistent theory}
\label{sec:TBG_model}

Before illustrating our numerical results, we would like to  briefly summarize the single-particle band model we have used to describe TBG and the self-consistent Hartree procedure we have carried out to deal with the important ground-state charge density inhomogeneities displayed by TBG.

\subsection{TBG bare-band model}
\label{subsect:TBG bare-band model}

The continuum model of TBG adopted in this work is the same as the one used in Ref.~\cite{novelli}, which was first derived in Refs.~\cite{bistritzer_pnas_2011} and~\cite{koshino_prx_2018}.

Layer, sublattice, spin, and valley are the four discrete degrees of freedom characterizing single-electron states in TBG. We can take into account valley and spin degrees of freedom by a degeneracy factor $g = 4 = g_{\rm v}g_{\rm s}$, where the spin-degeneracy factor $g_{\rm s}=2$ has been introduced earlier. The single-particle Hamiltonian of TBG is written in the layer/sublattice basis $\{|1A\rangle, |1B\rangle, |2A\rangle, |2B\rangle\}$ as:
\begin{equation}\label{eqn:continuumtot}
    \hat{{\cal H}}_0 = \left(\begin{array}{cc}
       \hat{{\cal H}}^{(1)}  & \hat{U} \\
        \hat{U}^\dagger & \hat{{\cal H}}^{(2)}
    \end{array}\right)~.
\end{equation}
The state $|\ell \tau\rangle$ refers to layer $\ell = 1, 2$ and sublattice index $\tau = A, B$, $\hat{\cal H}^{(\ell)}$ is the intra-layer Hamiltonian for layer $\ell$, and the operator $\hat{U}$ describes inter-layer tunneling. For
small twist angles, the moir\'e length scale $\sim a/\theta$ is much larger than the lattice parameter $a$ of single-layer graphene. This allows us to replace 
$\hat{\cal H}^{(\ell)}$ by its $\bm{k}\cdot \bm{p}$ massless Dirac fermion limit. This low-energy expansion is done around one of the single layer valleys, $K^{(\ell)}/K^\prime{}^{(\ell)}$:
\begin{equation}\label{eqn:continuumintra}
    \hat{{\cal H}}^{(\ell)} = v_{\rm D}\left[{\cal R}_\ell(\theta/2)(\hat{\bm{p}} \mp \hbar \bm{K}_{\ell})\right]\cdot(\pm\sigma_x,-\sigma_y)~.
\end{equation}
Here, $(\pm\sigma_x,-\sigma_y)$ is a vector of $2\times2$ Pauli matrices (the $\pm$ sign referring to the $K$ and $K^\prime$ valleys, respectively), $\hat{\bm p}$ is the momentum operator, $v_{\rm D} = \sqrt{3}|t|a/(2\hbar) \sim 1 \times
10^6$m/s is the Fermi velocity of single-layer graphene, $|t| = 2.78~{\rm eV}$ being the usual single-particle nearest-neighbor hopping. The vector $\bm{K}_\ell$ appearing in Eq.~\eqref{eqn:continuumintra} is the position of single layer graphene’s valley $K^{(\ell)}$ measured from the moir\'{e} BZ center $\Gamma$ (Fig.~\ref{fig:sketch+mBZ}~(b)):
\begin{equation}\label{k_point}
    \bm{K}_{1,2} = \frac{8\pi}{3 a}\sin{\left(\frac{\theta}{2}\right)}\left(-\frac{\sqrt{3}}{2},\pm\frac{1}{2}\right)~.
\end{equation}
The rotation matrix ${\cal R}_\ell(\theta/2)$ appearing in \eqref{eqn:continuumintra} is given by:
\begin{eqnarray}
    {\cal R}_{\ell=1,2}\left(\theta/2\right) &=& \cos(\mp\theta/2) \mathbb{I}_{2\times2} - i\sin(\mp\theta/2)\sigma_y \nonumber\\
    &=&\left(\begin{array}{cc}
        \cos\theta/2 & \pm\sin\theta/2 \\
        \mp\sin\theta/2 & \cos\theta/2
    \end{array}\right)~.
\end{eqnarray}
The convention adopted is such that $\theta_{\ell=1} = -\theta/2$ and $\theta_{\ell=2} = \theta/2$. The longitudinal displacement between the two layers is taken as zero in order to obtain the AB-Bernal stacking configuration for $\theta=0$.

The $\hat{U}$ operator describes inter-layer hopping and is given by:
\begin{eqnarray}\label{eqn:interlayeroperator}
    \hat{U} &=& \left(\begin{array}{cc}
       u_0  & u_1 \\
       u_1  & u_0
    \end{array}\right) + e^{-i\frac{2\pi}{3}+i\bm{G}_1\cdot\hat{\bm{r}}} \left(\begin{array}{cc}
       u_0  & u_1 e^{i\frac{2\pi}{3}}\\
       u_1e^{-i\frac{2\pi}{3}}  & u_0
    \end{array}\right) +\nonumber\\ &+&e^{i\frac{2\pi}{3}+i\bm{G}_2\cdot\hat{\bm{r}}} \left(\begin{array}{cc}
       u_0  & u_1 e^{-i\frac{2\pi}{3}}\\
       u_1e^{i\frac{2\pi}{3}}  & u_0
    \end{array}\right)~,
\end{eqnarray}
where
\begin{equation}\label{eqn:reciprocal_lattice_basis}
    \bm{G}_{1,2} = \frac{8\pi}{\sqrt{3} a} \sin\left(\frac{\theta}{2}\right)\left(\pm \frac{1}{2},\frac{\sqrt{3}}{2} \right)~,
\end{equation}
and $u_0$ ($u_1$) are the intra-sublattice  (inter-sublattice) hopping parameters. In general $u_0\neq u_1$. The difference between these two parameters can, in fact, take into account the lattice corrugation of TBG samples~\cite{koshino_prx_2018,lucignano_prb_2019,carr,carr_rev_2020}. The intra- and inter-sublattice hopping energies might also be affected in value by possible stresses induced on the TBG sheet during the production phase. Recently~\cite{hesp} it has been shown experimentally that the difference between the intra- and inter-sublattice hopping parameters is in the range of $u_1-u_0 \sim 30$-$60~{\rm meV}$. In this work, we take $u_1= 97.5~{\rm meV}$ and $u_0=79.7~{\rm meV}$. With this choice, we have $u_1-u_0 \approx 20~{\rm meV}$ and the dimensionless parameter $u_0/u_1 \sim 0.8$ takes correctly into account relaxation effects~\cite{koshino_prx_2018}.
Within the continuum model described by the single-particle Hamiltonian in Eq.~\eqref{eqn:continuumtot}, we can construct the projector operators onto the $i$-th layer $\hat{\Pi}^{(i)}$ by making explicit their action on the basis $|\ell\tau\rangle$:
\begin{equation}\label{eqn:projector_def}
    \hat{\Pi}^{(i)}|\ell\tau\rangle = |i \tau\rangle~.
\end{equation}
In particular their matrix form is given explicitly by:
\begin{equation}
    \hat{\Pi}^{(1)} = \left(\begin{array}{cc}
        \hat{\mathbb{I}}_{2\times2} & 0 \\
        0 & 0
    \end{array}\right)~,
\end{equation}
\begin{equation}
    \hat{\Pi}^{(2)} = \left(\begin{array}{cc}
        0 & 0 \\
        0 & \hat{\mathbb{I}}_{2\times2}
    \end{array}\right)~,
\end{equation}
where $\hat{\mathbb{I}}_{2\times2}$ is the identity operator acting on the sublattice index.

The chemical potential $\mu$ in Eq.~(\ref{eq:Fermi-Dirac}) can be calculated by enforcing, as usual, particle-number conservation:
\begin{equation}\label{eqn:density}
    n = \delta n + n_0 = g \sum_\lambda \int \frac{d^2\bm{k}}{(2\pi)^2} f_{\bm{k},\lambda}^{\rm reg}(\mu)~.
\end{equation}
Here, $n_0$ is the total electron density at the charge neutrality point (CNP) and $\delta n$ is the electron density measured from the CNP. We stress that a {\it regularized} Fermi-Dirac distribution function $f_{\bm{k},\lambda}^{\rm reg}$ appears in Eq.~\eqref{eqn:density}. Indeed, since we are dealing with a continuum model, the number of bands is formally infinite below and above the CNP. In order to regularize the Dirac sea below the CNP, one needs to introduce the regularized Fermi-Dirac distribution function defined as following:
\begin{align}
    f_{\bm{k},\lambda}^{\rm reg}(\mu) &\equiv f^{\rm reg}(\epsilon_{\bm{k},\lambda}-\mu)  =\nonumber\\
&=f(\epsilon_{\bm{k},\lambda}-\mu) - \Theta(\epsilon_{\rm CNP} - \epsilon_{\bm{k},\lambda})~,
\end{align}
where $\Theta(x)$ is the Heaviside step-function and $\epsilon_{\rm CNP}$ is the energy of the CNP.

With these conventions, the filling factor $\nu$ is defined by:
\begin{equation}\label{eqn:filling_factor}
    \nu \equiv \Omega_{\rm u.c.} \delta n~,
\end{equation}
where $\Omega_{\rm u.c.} = \frac{\sqrt{3}}{2}\left[\frac{a}{2\sin{(\theta/2)}}\right]^2$ is the area of the moir\'e unit cell. With this definition of the filling factor, one has $|\nu| < 4$ when the chemical potential is within the flat bands, at low temperatures.

\subsection{Hartree self-consistency}
\label{sect:TBG_hartree}

Inhomogeneities in the ground-state charge density distribution of TBG create an inhomogeneous electrical potential that depends on the filling factor. To capture this effect, we need to add the so-called Hartree contribution $\hat{V}_{\rm H}$ to the bare TBG Hamiltonian $\hat{\cal H}_0$~\cite{GiulianiVignale,novelli,guinea_2018}:
\begin{equation}\label{eq:Hartree_Hamiltonian}
    \hat{\cal{H}} = \hat{\cal{H}}_0 + \hat{V}_{\rm H}[n_{\bm G}]~,
\end{equation}
where
\begin{equation}\label{eq:Hartree_potential_explicit}
    \hat{V}_{\rm H}[n_{\bm G}] = \mathbb{I}_{4\times4}\sum_{\bm{G}\neq\bm{0}}\frac{2\pi e^2}{\bar{\varepsilon}|\bm{G}|}n_{\bm G}e^{i\bm{G}\cdot\hat{\bm r}}~.
\end{equation}
Here, $\bar{\varepsilon}\equiv (\varepsilon_1 + \varepsilon_3)/2$ , $n_{\bm G}$ is the Fourier component of the ground-state electron density corresponding to the reciprocal lattice vector $\bm{G}$, and the identity matrix $\mathbb{I}_{4\times4}$ is expressed in the same basis of states of the Hamiltonian, namely $\{|1A\rangle, |1B\rangle, |2A\rangle, |2B\rangle\}$. 

The problem
posed by Eqs.~(\ref{eq:Hartree_Hamiltonian})-(\ref{eq:Hartree_potential_explicit}) needs to be solved self-consistently,
i.e., one needs to solve the Hartree equation
\begin{equation}
    \left(\hat{\cal{H}}_0 + \hat{V}_{\rm H}[n_{\bm G}]\right)|\bm{k},\lambda\rangle = \epsilon_{\bm{k},\lambda}|\bm{k},\lambda\rangle~,
\end{equation}
together with the closure:
\begin{equation}
    n_{\bm G} = g\sum_{\lambda}\int \frac{d^2\bm{k}}{(2\pi)^2} f_{\bm{k},\lambda}^{\rm reg}\langle \bm{k},\lambda|e^{-i\bm{G}\cdot\hat{\bm r}}|\bm{k},\lambda\rangle~.
\end{equation}
Note that, due to the real-space representation~\eqref{eq:single-particle-Bloch-state} of the Bloch eigenstates, we have:
\begin{align}
&\langle\bm{k}, \lambda|e^{-i\bm{G}\cdot\hat{\bm{r}}}|\bm{k},\lambda\rangle =\nonumber\\
&=\frac{1}{S} \sum_{\bm{K},\bm{K}^\prime} u_{\bm{K}}^{\dagger}(\bm{k},\lambda) u_{\bm{K}^\prime}(\bm{k},\lambda)\int d^2\bm{r} e^{-i(\bm{G}+\bm{K}+\bm{k}-\bm{K}^\prime-\bm{k})\cdot\bm{r}}\nonumber\\
&= \sum_{\bm{K},\bm{K}^\prime} u_{\bm{K}}^{\dagger}(\bm{k},\lambda) u_{\bm{K}^\prime}(\bm{k},\lambda)\delta_{\bm{G+\bm{K},\bm{K}^\prime}}\nonumber\\
&= \sum_{\bm{K}} u_{\bm{K}}^{\dagger}(\bm{k},\lambda) u_{\bm{K}+\bm{G}}(\bm{k},\lambda)~.
\end{align}

Once the self-consistent problem has been solved, the Hartree eigenstates $|{\bm k}, \lambda\rangle$ and eigenvalues $\epsilon_{{\bm k}, \lambda}$ can be used in order to calculate the so-called Hartree density-density response~\cite{GiulianiVignale} matrix. This is simply obtained by using Eq.~\eqref{eq:density-density_xstal_layer}, with the understanding that the two quantities $|{\bm k}, \lambda\rangle$ and  $\epsilon_{{\bm k}, \lambda}$ in there need to be interpreted as self-consistently calculated Hartree quantities rather than single-particle, bare quantities.

\section{Numerical Results}\label{sec:numerical}
In this Section we present our main numerical results obtained with the theory outlined above. For the sake of definiteness, we set  $\varepsilon_1=\varepsilon_3=4.9$,  $\varepsilon_2=1$, and $T=5~{\rm K}$. 

The dielectric tensor and hence the loss function are obtained by using the calculated Hartree self-consistent bands and corresponding Bloch states. These calculations take into account the role of static screening in reshaping the electronic bands and redistributing in space the carrier density. The Hartree self-consistency effect on plasmons is more important at small twist angles, since in this regime the system displays larger charge inhomogeneities~\cite{novelli}. This is true also for the LFEs. 

\begin{figure}
\includegraphics[width=1\columnwidth]{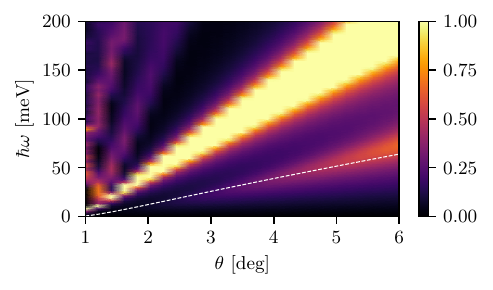}%
\caption{(Color online) The energy loss function ${\cal L}({\bm q}, \omega)$ of TBG is plotted as a function of the twist angle $\theta$ and frequency $\omega$. Results in this plot have been obtained by keeping fixed the wave number $q$ and filling factor $\nu$, i.e.~ $q = q_{\theta} \equiv 2|\bm{K}_{1,2}|/31$ (see main text) and $\nu=+1$. Bright bands correspond to plasmons peaks.  The white dashed line indicates the upper edge of the particle-hole continuum, i.e.~$\omega= v^{\star}(\theta)q_\theta$, $v^{\star}(\theta)$ being the reduced Fermi velocity, above which collective modes are well defined. The low-energy acoustic plasmon mode, which ``tracks'' the upper edge of the particle-hole continuum, disappears for $\theta \lesssim 2^{\degree}$.}
    \label{fig:loss_vs_theta}
\end{figure}
\begin{figure*}
\begin{tabular}{lll}
    \begin{overpic}[width=0.5\textwidth]{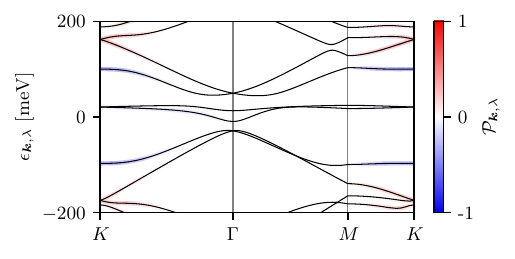}
    \put(0,48){(a)}
    \end{overpic} & \begin{overpic}[width=0.5\textwidth]{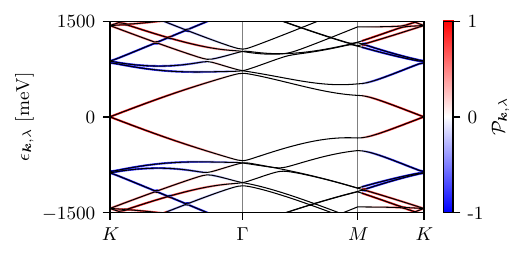}
    \put(0,48){(b)}
    \end{overpic}
\end{tabular}
\caption{(Color online) Layer polarization ${\cal P}_{\bm{k},\lambda}$ of the Hartree self-consistent eigenstates , superimposed on TBG energy bands calculated with Hartree self-consistency at filling factor $\nu=+1$. Panel (a) $\theta = 1.05^{\degree}$. Panel (b) $\theta = 5^{\degree}$. At lower angles the Hamiltonian eigenstates are less layer polarized, resulting in more hybridization and the suppression of the bi-layer acoustic plasmon mode. Bands are calculated at the $K^\prime$ valley.}
\label{fig:layer_polarization}
\end{figure*}
Fig.~\ref{fig:acoustic_appearence} shows the TBG loss function for filling factor $\nu=+1$ and two values of the twist angle $\theta$, i.e.~$\theta = 1.05^{\degree}$ in panel (a) and $\theta = 5^{\degree}$ in panel (b). This filling factor corresponds to a carrier density $n=0.64\times10^{12}$~cm$^{-2}$ for $\theta = 1.05^{\degree}$ and $n=1.5\times10^{13}$~cm$^{-2}$ for $\theta = 5^{\degree}$. Chemical potential values have been given in the caption of Fig.~\ref{fig:acoustic_appearence}. Close to the magic angle, Fig.~\ref{fig:acoustic_appearence}(a), flat bands centered at the CNP and separated by an energy gap from the higher-energy bands, lead to intrinsically undamped slow plasmons~\cite{levitov}. We clearly see this in Fig.~\ref{fig:acoustic_appearence}(a), where a narrow, almost dispersion-less plasmon is present at energies on the order of $\sim 20~{\rm meV}$. In general, we find that, at small twist angles, TBG hosts a standard intra-band COM plasmon with a $\omega_{\rm COM}(q)\propto\sqrt{q}$ dispersion in the long-wavelength limit. No sign of other collective modes is seen at small values of $\theta$, neither gapless~\cite{santoro_giuliani,profumo} nor gapped~\cite{Dassarma_prl_1998,Abedinpour_prl_2007}---further results are reported in Section II of Ref.~\cite{SM}. 

This is not the case for larger values of the twist angle, as seen for example in Fig.~\ref{fig:acoustic_appearence}(b) for $\theta = 5^{\degree}$. For this value of the twist angle, an acoustic plasmon is clearly visible. This mode lies just above the upper edge of the particle-hole continuum (Section I of Ref.~\cite{SM}), which is identified by the line $\hbar\omega_\theta(q) = \hbar v^{\star}_\theta q$,  $v^{\star}_\theta$ being the reduced Fermi velocity of the TBG Dirac cones~\cite{bistritzer_prb_2010}:
\begin{equation}\label{eqn:normalized_fermi_velocity}
    v^{\star}_{\theta} = v_{\rm D} \frac{1 - 3\alpha^2(\theta)}{1+6\alpha^2(\theta)}~,
\end{equation}
$\alpha(\theta) = u_1\left[\frac{8\pi}{\sqrt{3} a}\hbar v_{\rm D} \sin\left(\frac{\theta}{2}\right)\right]^{-1}$ being a dimensionless parameter that depends on the twist angle (the parameters $v_{\rm D}$ and $u_1$ have been introduced in Sect.~\ref{subsect:TBG bare-band model}). For $\theta = 5^{\degree}$, the Fermi velocity~\eqref{eqn:normalized_fermi_velocity} is $v_{\theta}^* \approx 7.99\times10^{5}$~m/s, while the acoustic plasmon velocity in Fig.~\ref{fig:acoustic_appearence}(b) is $c_{\rm s} \approx 8.43\times10^{5}$~m/s. For the sake of comparison, we note that the acoustic plasmon velocity in two (tunnel-decoupled but Coulomb-coupled) graphene layers at a distance $d=0.3$~nm is $c_{\rm s}\approx1.2\times10^{6}$~m/s (and at the same density $n=1.5\times10^{13}$~cm$^{-2}$)~\cite{profumo}. A reduced single-particle Fermi velocity in TBG leads to slower acoustic plasmons with respect to other graphene-related systems~\cite{profumo}. A plot illustrating the dependence of $c_{\rm s}$ on $\theta$ is reported in Section I of Ref.~\cite{SM}.

(Further numerical results are reported in Section II of Ref.~\cite{SM}---where the plasmon dispersion relation obtained with the inclusion of the layer-pseudospin degree of freedom and LFEs is compared with that obtained by neglecting the latter---and Section III of Ref.~\cite{SM}--- where the dependence on the filling factor $\nu$ is discussed, for various twist angles.
In Section II of Ref.~\cite{SM}, we note that the introduction of LFEs leads to a blue shift in the energy of the plasmon modes around the edge of the moir\'{e} BZ, as already found out in other systems~\cite{cudazzo,rubio,lewandowski}. This effect is even more pronounced at small twist angles. 
In Section III of Ref.~\cite{SM}, we observe, for a fixed value of $\theta$, a weak dependence on $\nu$. The impact of an applied perpendicular electric field and heterostrain on the plasmonic spectrum of TBG are discussed in Sects.~IV and~V of Ref.~\cite{SM}, respectively.)

Fig.~\ref{fig:loss_vs_theta} shows the loss function ${\cal L}({\bm q}, \omega)$ as a function of the twist angle $\theta$ and frequency $\omega$. Results in this figure have been obtained by setting $q=q_\theta \equiv \xi |\bm{K}_{1,2}|$, where $|\bm{K}_{1}|= |{\bm K}_{2}|$ is the modulus of the $\theta$-depending vector linking $\Gamma$ to $K$ in the moir\'{e} BZ---see Eq.~\eqref{k_point}---and $\xi=2/31<1$. The brightest feature in this figure corresponds to the usual COM plasmon while the lower-energy feature corresponds to the acoustic plasmon. At  twist angles $\theta \lesssim 2^{\degree}$, the acoustic plasmon branch disappears. We conclude that, at small twist angles, low energies, and long wavelengths, TBG behaves effectively as a single 2DES with an ordinary COM plasmon. A weakly-damped out-of-phase acoustic plasmon appears only for twist angles larger than $\theta \approx 2^{\degree}$. As discussed in Sect~\ref{sect:intro}, this mode is typical of weakly-coupled double layers, where two spatially-separated 2DESs interact only through the long-range Coulomb interaction~\cite{santoro_giuliani,profumo}. The gapless nature of the extra mode emerging for $\theta \gtrsim 2^{\degree}$ is reasonable  since the moir\'e potential that couples the two layers does not open a gap at the $K$/$K^\prime$ points (Dirac cones are protected by symmetry).

Despite the apparent similarity with spatially-separated 2DESs, acoustic plasmons in TBG offer a qualitative difference: in the latter system, they emerge only for sufficiently large values of $\theta$. In the former systems, instead, acoustic plasmons exist for all values of the macroscopic parameters, provided that the single-particle Fermi velocities in the two 2DESs are identical~\cite{santoro_giuliani,profumo}.

Regarding damping of the TBG acoustic plasmon, let us recall that the upper edge of the particle-hole continuum in TBG is given by:
\begin{equation}\label{eqn:particle_hole_theta}
    \hbar \omega_{\theta}(q_\theta) \equiv \hbar v^*_\theta q_\theta = \xi \frac{8\pi}{\sqrt{3}a} \hbar v_{\rm D} \frac{\sin^2(\theta/2) - 3\tilde{\alpha}^2}{\sin^2(\theta/2) + 6\tilde{\alpha}^2}  \sin(\theta/2)~,
\end{equation}
where $\tilde{\alpha} = \alpha(\theta)/\sin(\theta/2)$ and $\alpha(\theta)$ has been introduced above in Eq.~\eqref{eqn:normalized_fermi_velocity}.
If the plasmon dispersion lies {\it above} this threshold value, it is a well-defined (i.e.~long lived) mode (at least within the RPA). Since the wave vector $q$ is fixed at the value $q_\theta \equiv \xi |\bm{K}_{1,2}|$, the expression on the right hand side of Eq.~\eqref{eqn:particle_hole_theta} depends only on $\theta$ and is plotted in Fig.~\ref{fig:loss_vs_theta} (white dashed line) for small values of $\theta$ (up to $\theta=6^{\degree}$). We clearly see that, for sufficiently large values of $\theta$ (i.e.~$\theta\gtrsim 4^{\degree}$) the acoustic plasmon is a well-defined long-lived collective mode. 

In order to better understand the disappearance of the acoustic mode for $\theta \lesssim 2^{\degree}$, we have calculated the layer polarization ${\cal P}_{\bm{k},\lambda}$ of the TBG Hartree self-consistent eigenstates $| \bm{k},\lambda\rangle$. This quantity is defined as~\cite{De_Beule_PRR_2020}:
\begin{equation}
    {\cal P}_{\bm{k},\lambda} \equiv \langle \bm{k},\lambda| \hat{\Pi}^{(1)}| \bm{k},\lambda\rangle - \langle \bm{k},\lambda| \hat{\Pi}^{(2)}| \bm{k},\lambda\rangle~,
\end{equation}
where $\hat{\Pi}^{(i)}$ is the projector operator onto the $i$-th layer introduced in Sec.~\ref{sec:TBG_model}, Eq.~\eqref{eqn:projector_def}.
Fig.~\ref{fig:layer_polarization} shows the layer polarization (color bar) at the $K$ valley and for two values of the twist angle, i.e.~$\theta = 1.05^{\degree}$---panel (a)---and $\theta = 5^{\degree}$---panel (b). For the latter value of the twist angle, the polarization is  $|{\cal P}_{\bm{k},\lambda}|\approx 1$ for almost every value of the wave vector ${\bm k}$ and throughout all the bands. At $\theta = 1.05^{\degree}$, instead, we observe a very low layer polarization stemming from a strong inter-layer hybridization. It is this transition from high to low values of the layer polarization that, in our opinion, leads to the disappearance of the acoustic plasmon mode at twist angles $\theta \lesssim 2^{\degree}$.

\section{Summary and conclusions}
\label{sect:summary}
In this Article we have presented a theoretical study of the plasmonic response of twisted bilayer graphene as a function of the twist angle $\theta$. Our theory treats on equal footing four important effects, namely the layer degree of freedom, non-local effects in the density-density response function beyond the dynamical long-wavelength limit, Hartree self-consistency, and crystalline local field effects. 

We have found that at small values of the twist angle ($\theta \lesssim 2^{\degree}$) and in the low-energy long-wavelength limit, the 2D electron system in twisted bilayer graphene responds to a perturbation carrying wave vector $q$ and energy $\hbar \omega$ as a single entity, displaying a center-of-mass mode $\omega_{\rm COM}(q)\propto \sqrt{q}$. This is in agreement with all earlier studies~\cite{stauber_2013,stauber_2016,levitov,novelli}. As the twist angle increases, however, inter-layer tunneling decreases and the layer-pseudospin becomes a quasi-good quantum number. For $\theta \gtrsim 2^{\degree}$, the layer-pseudospin degree of freedom needs to be taken into account and the plasmonic spectrum of the system displays a qualitatively different behavior. In this case, indeed, a weakly-damped acoustic plasmon mode appears, akin to the acoustic plasmon of other parallel 2D electron systems of historical importance~\cite{das_sarma_1981,santoro_giuliani}.

In the future it will be interesting to feed our results to an Eliashberg theory~\cite{Marsiglio_AP_2020} of plasmon-mediated superconductivity in twisted bilayer graphene and to study the spatial distribution of chirality associated to this mode~\cite{Stauber_NanoLett_2020,Lin_PRL_2020}.

\begin{acknowledgments}
This work was supported by i) the European Union's Horizon 2020 research and innovation programme under grant agreement no.~881603 - GrapheneCore3 and the Marie Sklodowska-Curie grant agreement No.~873028 and ii) the Italian Minister of University and Research (MUR) under the ``Research projects of relevant national interest  - PRIN 2020''  - Project no.~2020JLZ52N, title ``Light-matter interactions and the collective behavior of quantum 2D materials (q-LIMA)''. P.J.H. acknowledges support by the Gordon and Betty Moore Foundation’s EPiQS Initiative through grant GBMF9463, the Fundacion Ramon Areces, and the ICFO Distinguished Visiting Professor program. F.H.L.K. acknowledges financial support from  the ERC TOPONANOP  (726001), the Government of Catalonia trough the SGR grant, the Spanish Ministry of Economy and Competitiveness through the Severo Ochoa Programme for Centres of Excellence in R\&D (Ref. SEV-2015-0522) and Explora Ciencia (Ref. FIS2017- 91599-EXP), Fundacio Cellex Barcelona, Generalitat de Catalunya through the CERCA program, the Mineco grant Plan Nacional (Ref. FIS2016-81044-P), and the Agency for Management of University and Research Grants (AGAUR) (Ref. 2017-SGR-1656).

 L.C. gratefully wishes to thank Matteo Ceccanti and Simone Cassandra for help with the graphics work.
\end{acknowledgments}
%

\clearpage 
\newpage

\setcounter{section}{0}
\setcounter{equation}{0}%
\setcounter{figure}{0}%
\setcounter{table}{0}%

\setcounter{page}{1}

\renewcommand{\thetable}{S\arabic{table}}
\renewcommand{\theequation}{S\arabic{equation}}
\renewcommand{\thefigure}{S\arabic{figure}}
\renewcommand{\bibnumfmt}[1]{[S#1]}
\renewcommand{\citenumfont}[1]{S#1}

\onecolumngrid

\begin{center}
\textbf{\Large Supplemental Material for:\\ ``Theory of intrinsic acoustic plasmons in twisted bilayer graphene''}
\bigskip

Lorenzo Cavicchi,$^{1}$
Iacopo Torre,$^{2,\,3}$
Pablo Jarillo-Herrero,$^{4}$
Frank H. L. Koppens,$^{3,\,5}$
Marco Polini$^{6,\,3}$

\bigskip

$^1$\!{\it Scuola Normale Superiore, Piazza dei Cavalieri 7, I-56126 Pisa,~Italy}

$^2$\!{\it Departament de F\'{i}sica, Universitat Polit\`{e}cnica de Catalunya, Campus Nord B4-B5, 08034 Barcelona,~Spain}

$^3$\!{\it ICFO-Institut de Ci\`{e}ncies Fot\`{o}niques, The Barcelona Institute of Science and Technology, Av. Carl Friedrich Gauss 3, 08860 Castelldefels (Barcelona),~Spain}

$^4$\!{\it Department of Physics, Massachusetts Institute of Technology, Cambridge, Massachusetts,~USA}

$^5$\!{\it ICREA-Instituci\'{o} Catalana de Recerca i Estudis Avan\c{c}ats, Passeig de Llu\'{i}s Companys 23, 08010 Barcelona,~Spain}

$^6$\!{\it Dipartimento di Fisica dell'Universit\`a di Pisa, Largo Bruno Pontecorvo 3, I-56127 Pisa,~Italy}

\bigskip

In this Supplemental Material we present more numerical results for the energy loss function of TBG. We discuss further results concerning: i) acoustic plasmons and the particle-hole continuum; ii) the impact of LFEs on the plasmonic spectrum; iii) the robustness of the plasmonic spectrum with respect to changes in the filling factor; iv) effects of a static, perpendicular electric field; v) heterostrain effects on plasmons.
 
\end{center}

\onecolumngrid

\appendix
%
%
\setcounter{page}{1}
\section*{Section I: Acoustic plasmons and the TBG particle-hole continuum}\label{app:PH-continuum}
\begin{figure}[h!]
\centering
\begin{tabular}{lll}
    \begin{overpic}[width=0.5\textwidth]{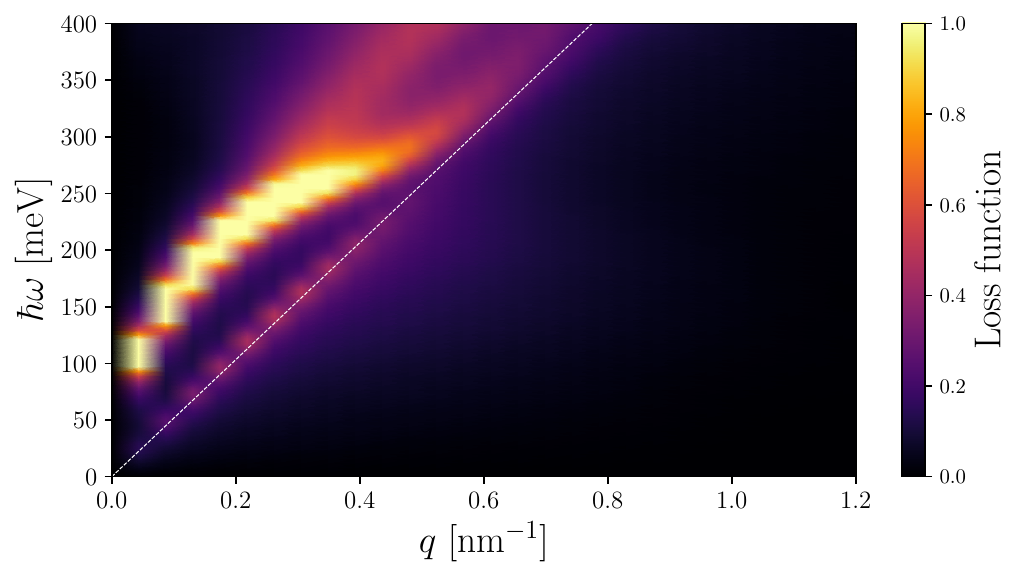}%
    \put(0,54){(a)}
    \end{overpic} & \begin{overpic}[width=0.5\textwidth]{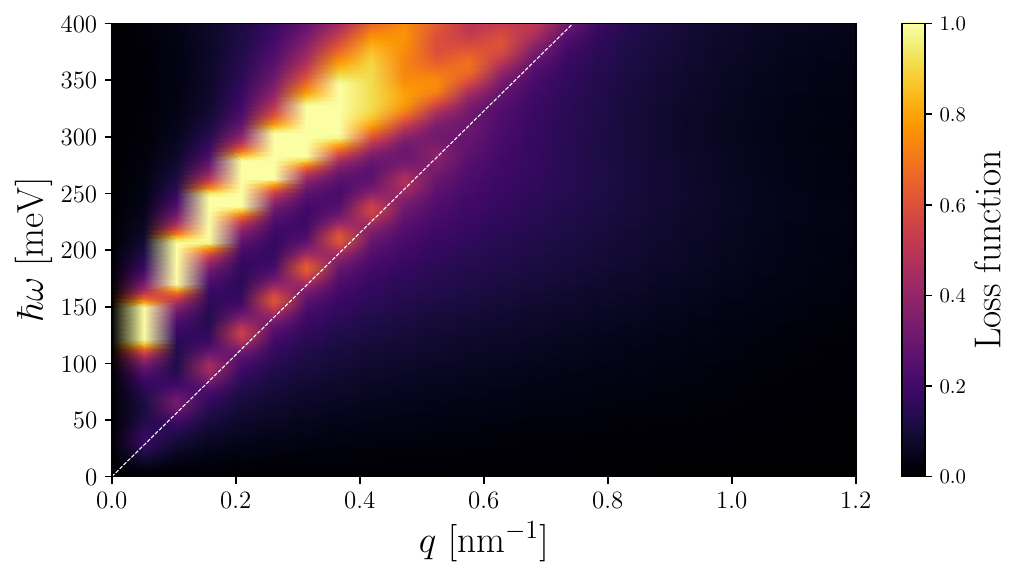}%
    \put(0,54){(b)}
    \end{overpic}
\end{tabular}
\caption{(Color online) The TBG energy loss function ${\cal L}({\bm q},\omega)$ is plotted as a function of ${\bm q}$ (along the $\Gamma-K$ high symmetry path) and $\omega$ for two values of the twist angle $\theta$: $\theta=5\degree$ in panel (a) and $\theta=6\degree$ in panel (b). Results in this plot refer to filling factor $\nu=+1$ and temperature 
$T= 5~{\rm K}$. The upper edge of the particle-hole continuum (which of course depends on $\theta$), i.e.~$\omega= v^{\star}_\theta q$, $v^{\star}_\theta$ being the reduced Fermi velocity (see Eq.~\eqref{eqn:normalized_fermi_velocity} in the main text), is represented by a thin white dashed line.
\label{fig:acoustic_appearence_appendix}}
\end{figure}

In this Section we show that the acoustic plasmon appearing in Fig.~\ref{fig:acoustic_appearence}(b) of the main text lies above the TBG particle-hole continuum (and it is therefore undamped). As discussed in the main text, the upper edge of such continuum is identified by $\hbar\omega_\theta = \hbar v^*_\theta q$. In Figs.~\ref{fig:acoustic_appearence_appendix}(a) and (b) we report the plasmon spectrum of TBG for $\theta = 5^\circ$ and $\theta = 6^\circ$, respectively. In each panel, the white dashed line represents the $\hbar\omega_\theta = \hbar v^*_\theta q$ line. We clearly see that, for both twist angles, the acoustic plasmon mode lies above the particle-hole continuum, although falls very close to it. 

\begin{figure}[h!]
\centering
\begin{overpic}[width=0.5\textwidth]{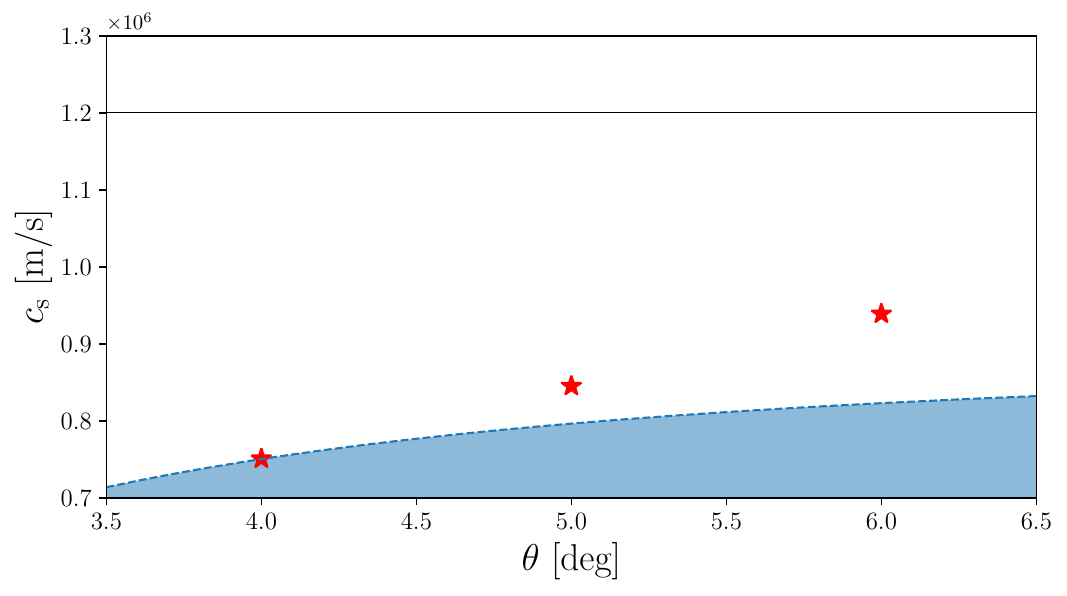}%
\put(58,11){{\color{white} Particle-hole continuum}}
\end{overpic}
\caption{(Color online) The sound velocity $c_{\rm s}$ (red stars) of the acoustic plasmon is plotted as a function of the twist angle $\theta$. Results in this plot have been obtained by setting $\nu = +1$ and $T=5$~K. The shaded region identifies the particle-hole continuum, whose upper edge coincides with $v^{\star}_\theta$ (see Eq.~\eqref{eqn:normalized_fermi_velocity} in the main text). The horizontal solid line represents the acoustic plasmon velocity in two spatially-separated graphene layers~\cite{profumo} at a distance $d=0.3$~nm and total electron density of $n=1.5\times10^{13}$~cm$^{-2}$. 
\label{fig:acoustic_appearence_appendix2}}
\end{figure}

A plot summarizing the dependence of $c_{\rm s}$ on $\theta$ is reported in Fig.~\ref{fig:acoustic_appearence_appendix2}. 

\section*{Section II: Impact of LFEs on the plasmonic spectrum}\label{app:numerical}
\begin{figure*}
\begin{tabular}{lll}
    \begin{overpic}[width=0.5\textwidth]{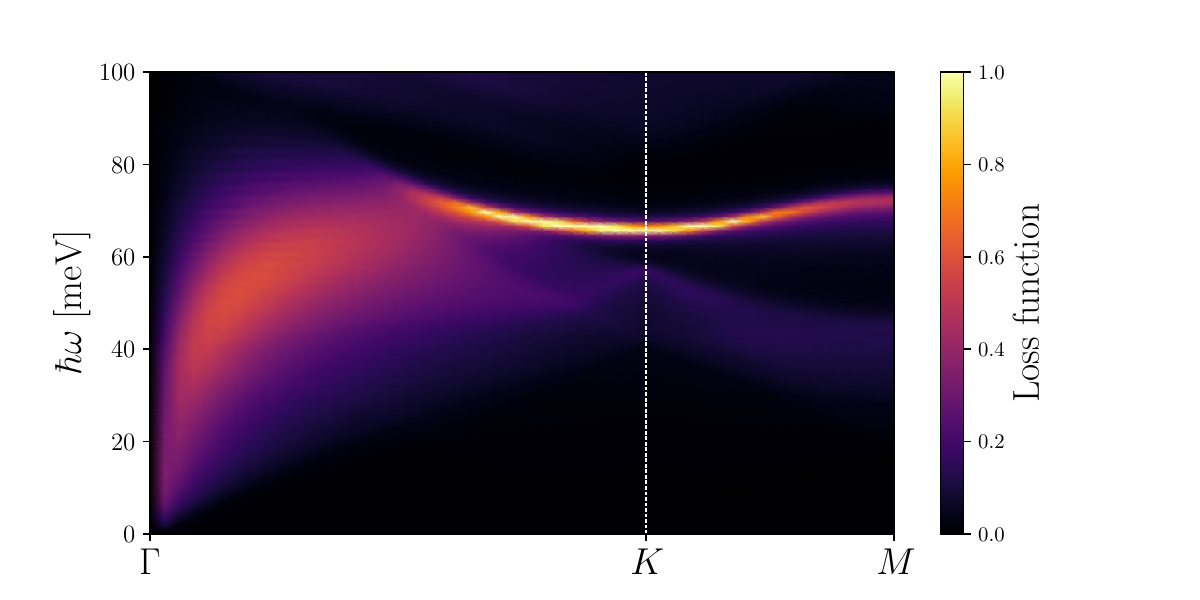}%
    \put(0,43){(a)}
    \end{overpic} & \begin{overpic}[width=0.5\textwidth]{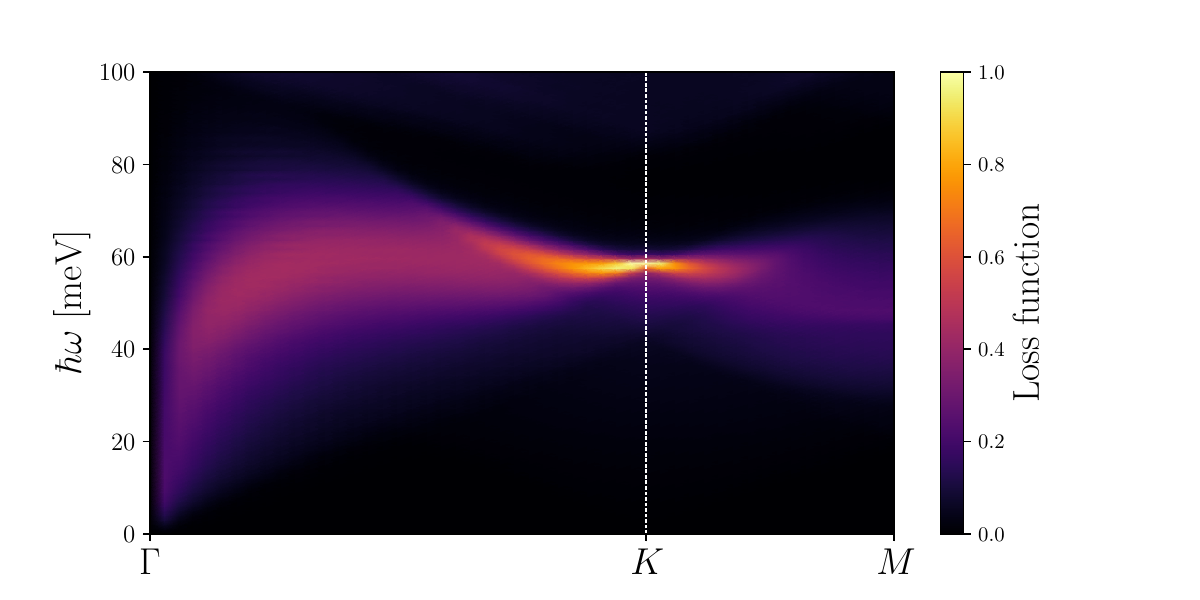}%
    \put(0,43){(b)}
    \end{overpic}\\
    \begin{overpic}[width=0.5\textwidth]{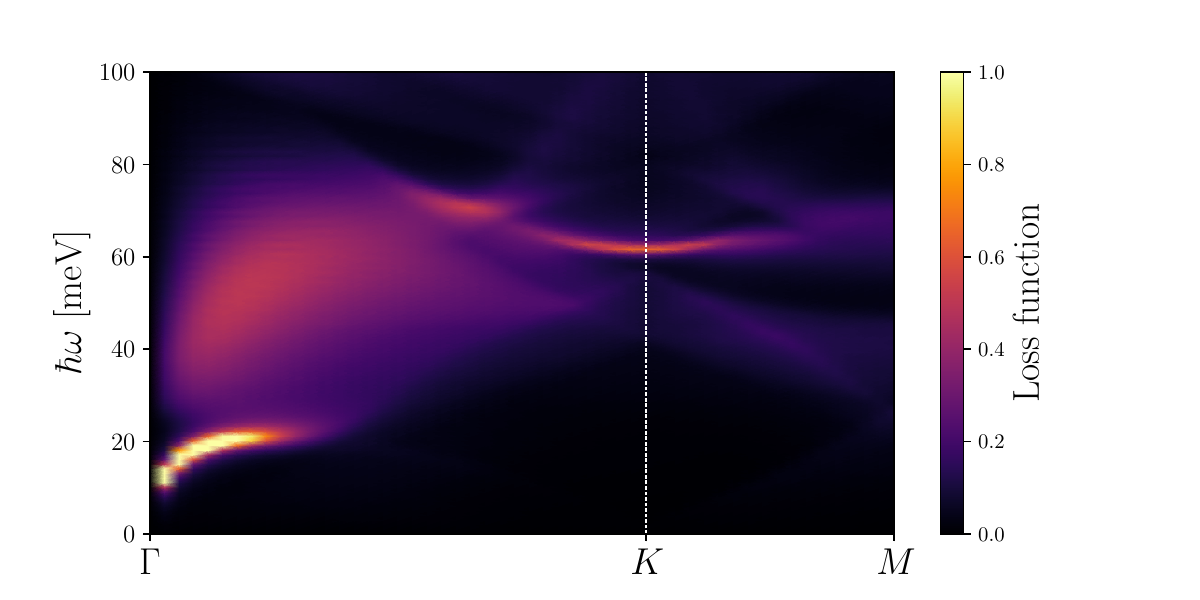}%
    \put(0,43){(c)}%
    \end{overpic} & \begin{overpic}[width=0.5\textwidth]{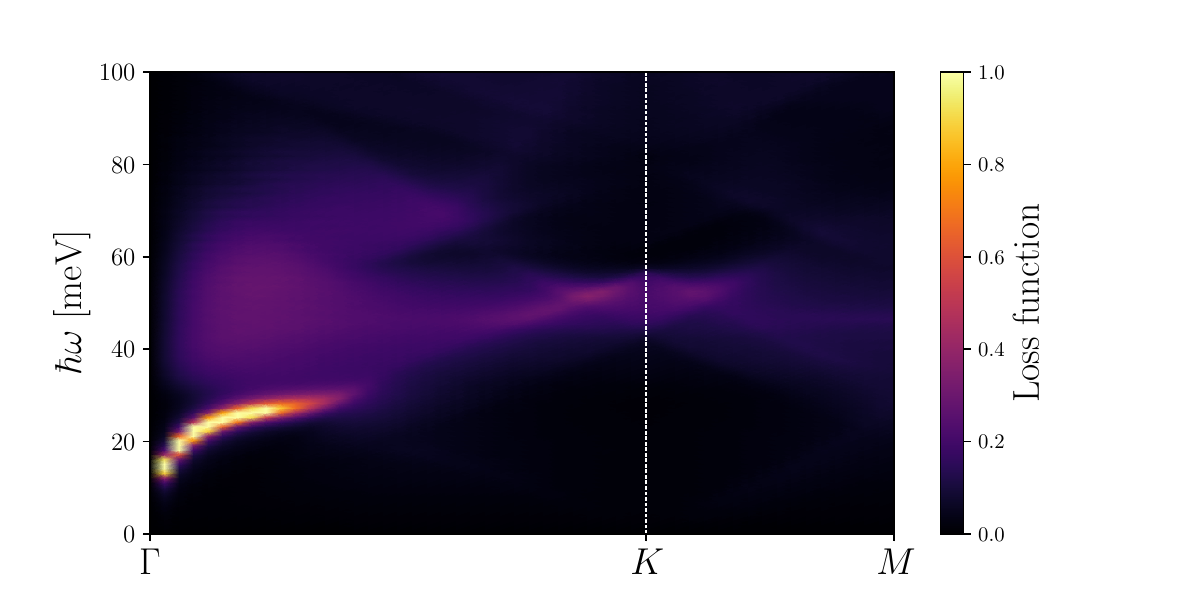}%
    \put(0,43){(d)}%
    \end{overpic}\\
    \begin{overpic}[width=0.5\textwidth]{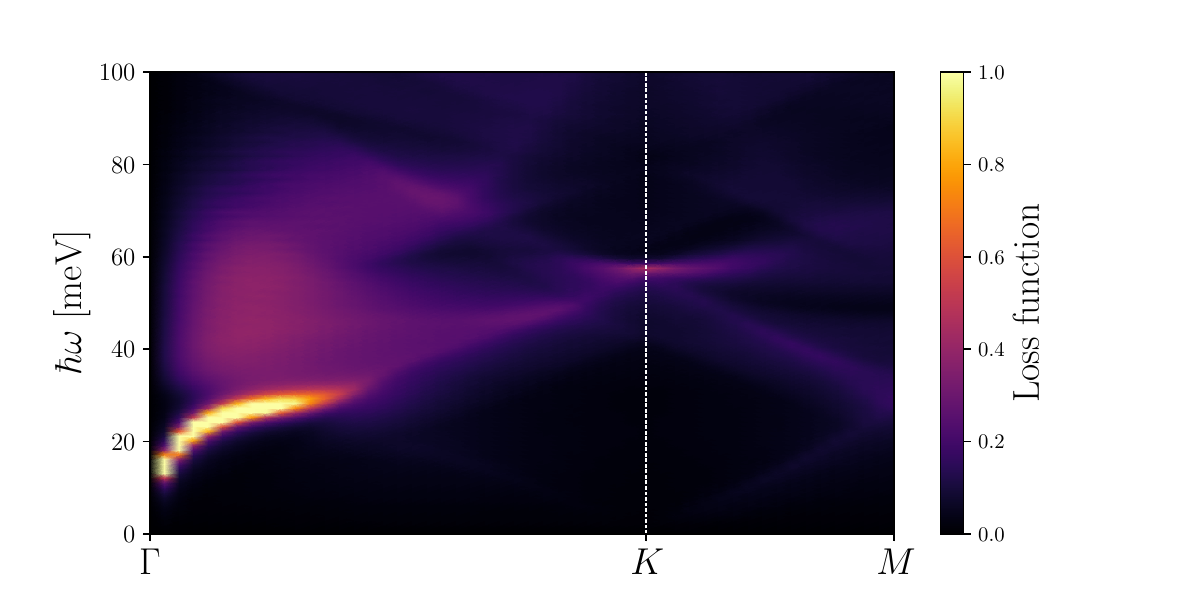}%
    \put(0,43){(e)}%
    \end{overpic} & \begin{overpic}[width=0.5\textwidth]{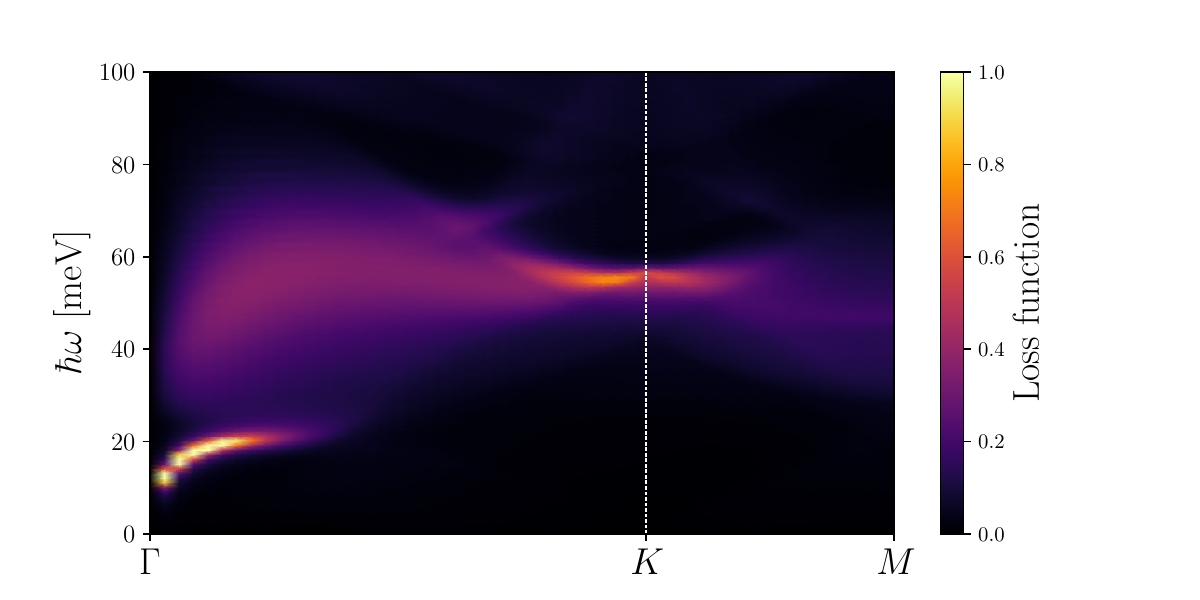}%
    \put(0,43){(f)}%
    \end{overpic}\\
\end{tabular}
\caption{(Color online) The TBG energy loss function ${\cal L}({\bm q},\omega)$ is plotted as a function of ${\bm q}$ and $\omega$ for $\theta=1.35^{\degree}$ and various filling factors $\nu$. Results shown in the panels on the left (right) column have been obtained by including (neglecting) LFEs. Panels (a)-(b): $\nu=0$. Panels (c)-(d): $\nu=+1$. Panels (e)-(f): $\nu=+2$. On the horizontal axis we report ${\bm q}$ along the high-symmetry path $\Gamma$-$K$-$M$ of the moir\'{e} BZ---see Fig.~\ref{fig:sketch+mBZ}(b) in the main text.\label{fig:loss_function_135}}
\end{figure*}

\begin{figure*}
\begin{tabular}{lll}
    \begin{overpic}[width=0.5\textwidth]{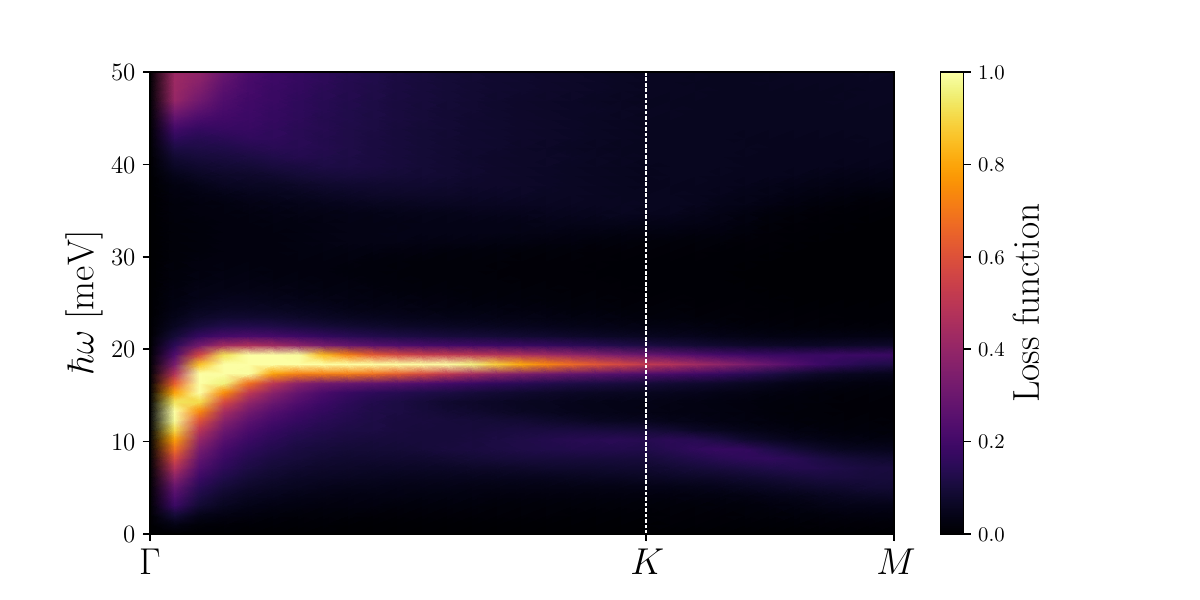}%
    \put(0,43){(a)}
    \end{overpic} & \begin{overpic}[width=0.5\textwidth]{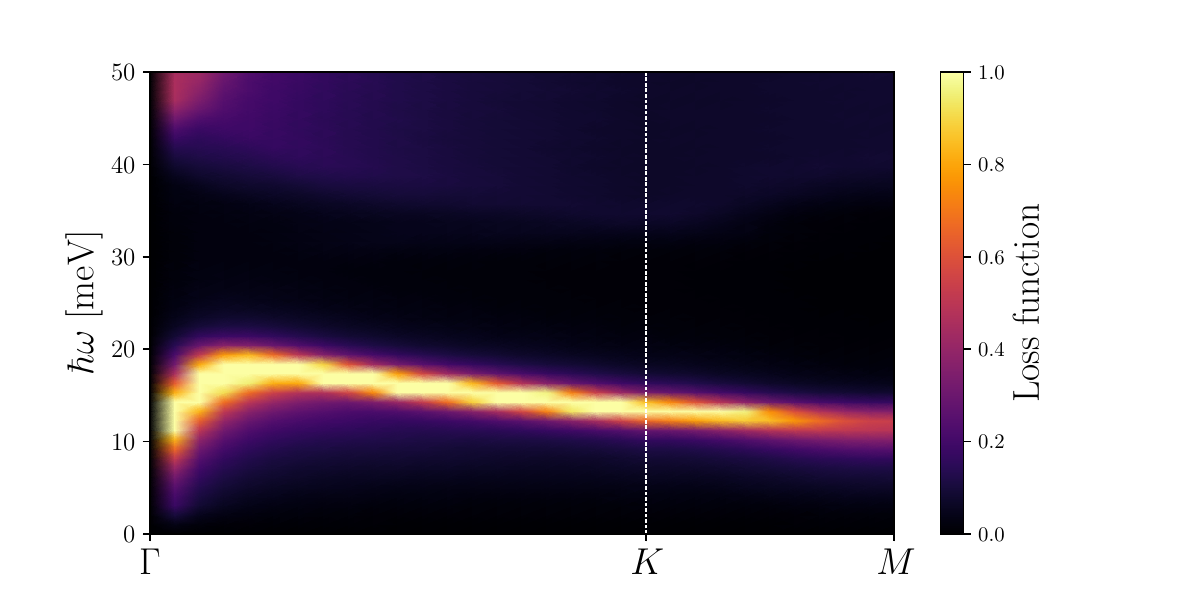}%
    \put(0,43){(b)}
    \end{overpic}\\
    \begin{overpic}[width=0.5\textwidth]{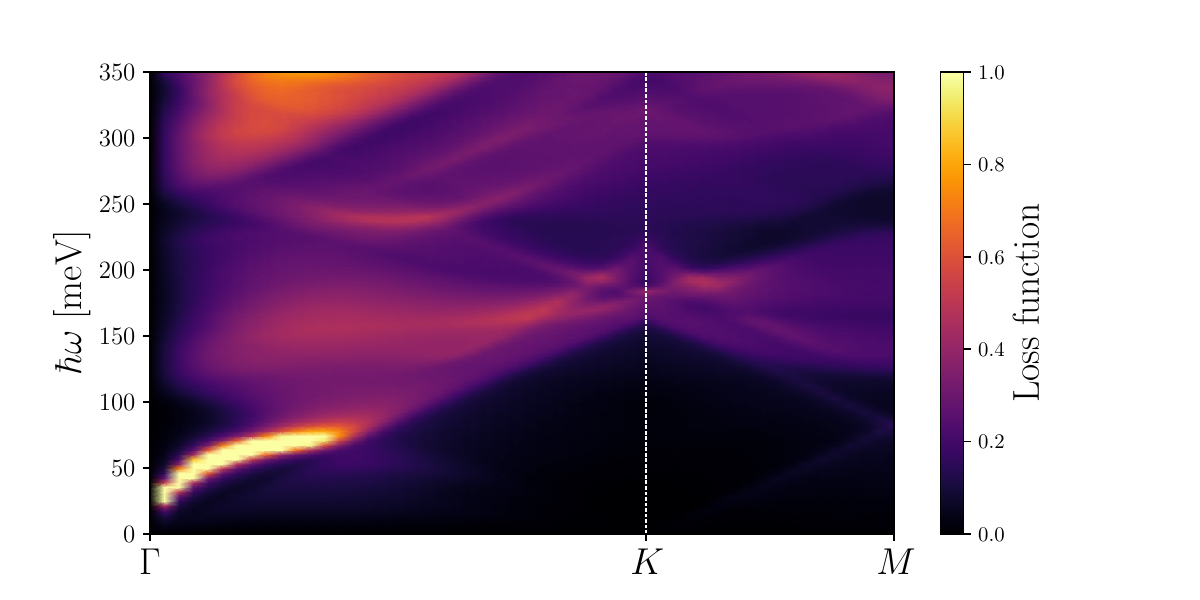}%
    \put(0,43){(c)}%
    \end{overpic} & \begin{overpic}[width=0.5\textwidth]{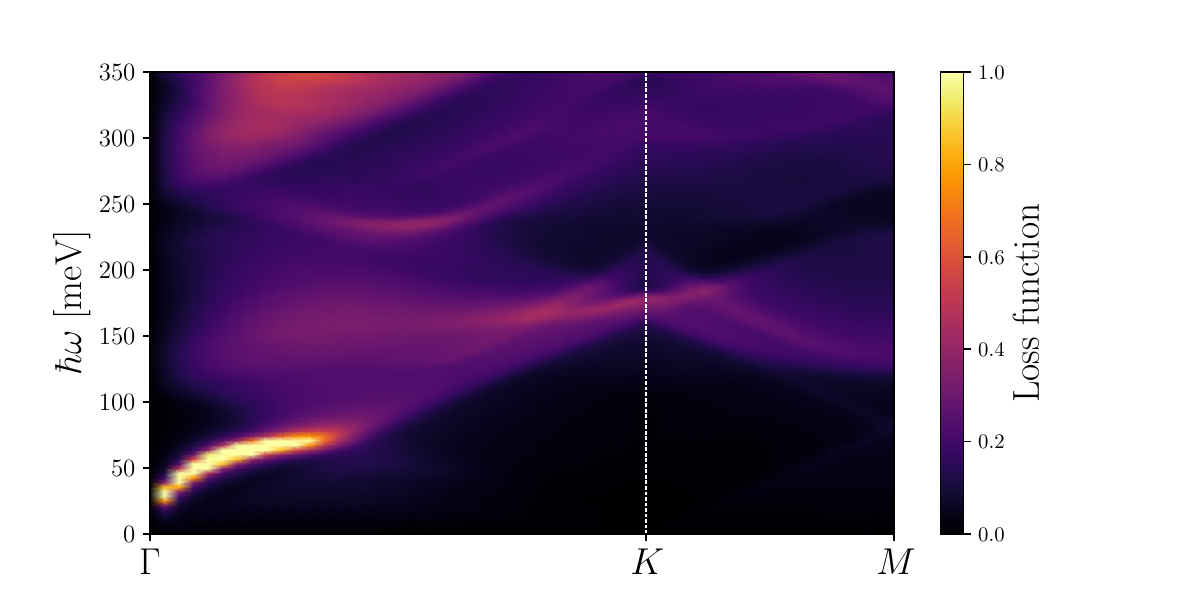}%
    \put(0,43){(d)}%
    \end{overpic}\\
    \begin{overpic}[width=0.5\textwidth]{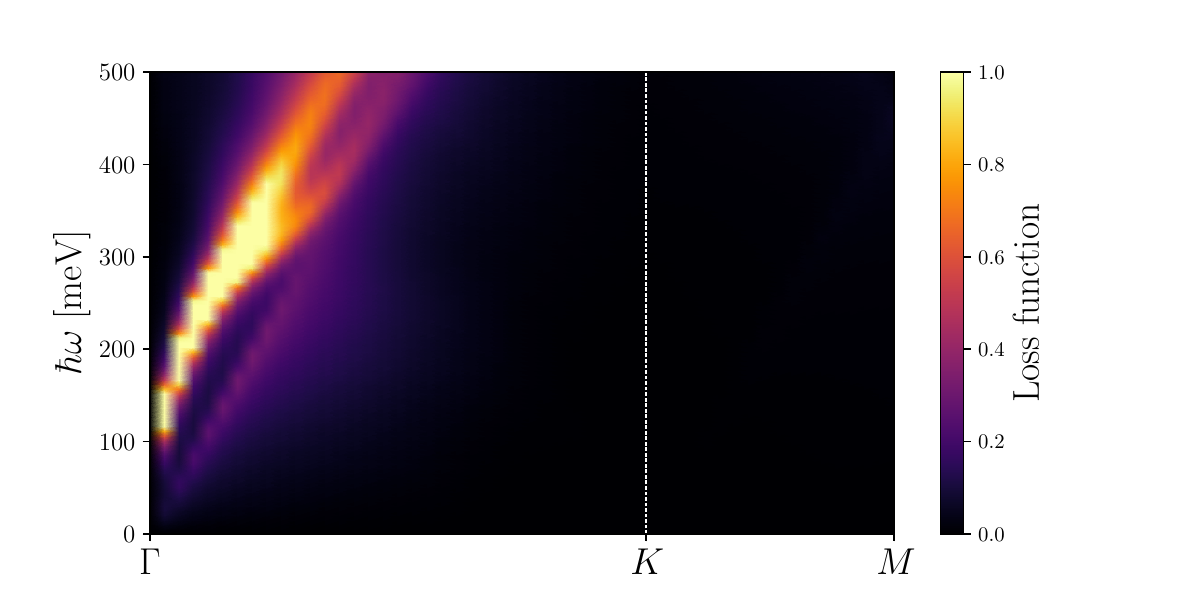}%
    \put(0,43){(e)}%
    \end{overpic} & \begin{overpic}[width=0.5\textwidth]{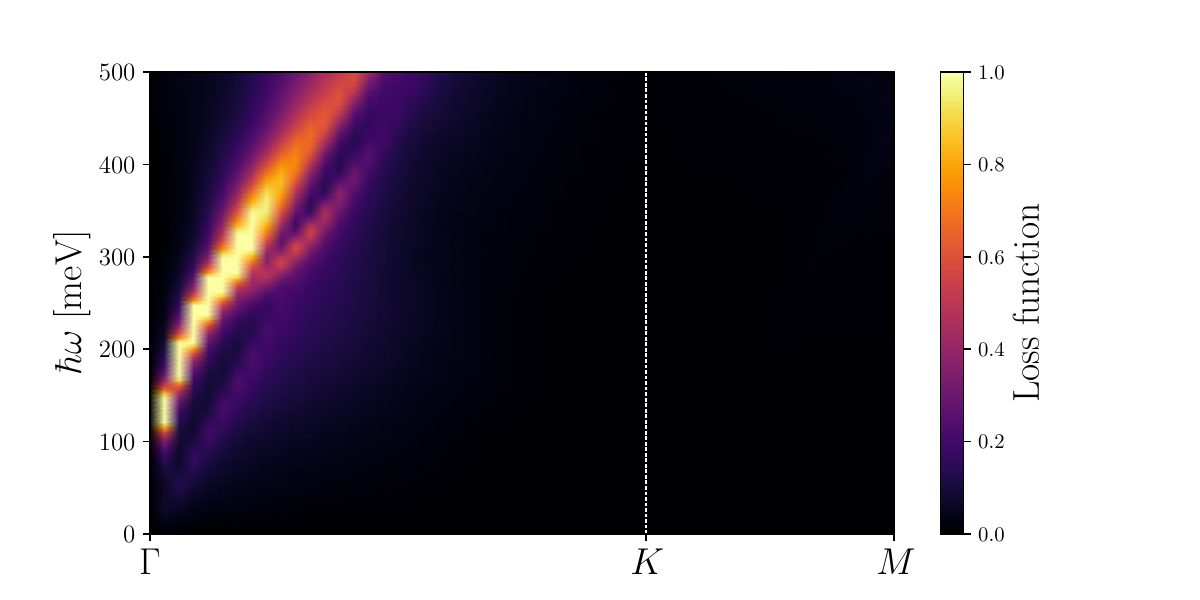}%
    \put(0,43){(f)}%
    \end{overpic}\\
\end{tabular}
\caption{(Color online) The TBG energy loss function ${\cal L}({\bm q}, \omega)$ is plotted as a function of ${\bm q}$ and $\omega$ for $\nu=+1$ and various values of the twist angle $\theta$. Results in the left column (i.e.~panels (a), (c), and (e)) have been obtained by including LFEs. Results in the right column (i.e.~panels (b), (d), and (f)) have been obtained by neglecting LFEs. Panels (a)-(b): $\theta=1.05^{\degree}$. Panels (c)-(d): $\theta=2^{\degree}$. Panels (e)-(f): $\theta=6^{\degree}$. On the horizontal axis we report ${\bm q}$ along the high-symmetry path $\Gamma$-$K$-$M$ of the moir\'{e} BZ---see Fig.~\ref{fig:sketch+mBZ}(b) in the main text.\label{fig:loss_function_angles}}
\end{figure*}

In this Section we discuss the role of LFEs on the plasmonic spectrum. Results presented in Figs.~\ref{fig:loss_function_135} and~\ref{fig:loss_function_angles} have been obtained by setting $\varepsilon_1=\varepsilon_3=4.9$,  $\varepsilon_2=1$, and $T=5~{\rm K}$. In order to isolate the impact of LFEs, we have deliberately neglected Hartree corrections in producing the data reported in Figs.~\ref{fig:loss_function_135} and~\ref{fig:loss_function_angles}.

Fig.~\ref{fig:loss_function_135} compares the energy loss function (for $\theta=1.35^{\degree}$ and various values of $\nu$) in the local (i.e.~ $\bm{G}=\bm{G}^\prime=\bm{0}$) approximation (panels in the right column) with that calculated by including LFEs (panels in the left columns). The impact of LFEs is most significant around the edges of the moir\'{e} BZ. This is especially true at low doping. 

A similar comparison is reported in Fig.~\ref{fig:loss_function_angles} where the filling factor is fixed at $\nu=+1$ while the twist angle is varied. Increasing the angle leads to a reduction of the importance of LFEs corrections. As emphasized in the main text, for $\theta=6^{\degree}$ we can clearly see the acoustic plasmon mode.

\section*{Section III: Robustness of the plamonic spectrum with respect to changes in the filling factor}
\label{app:filling}
In this Section we study the robustness of the acoustic plasmon with respect to changes in the filling factor (exploring, in particular, higher values of $\nu$, as compared to the main text). 

\begin{figure}[h!]
\begin{tabular}{ll}
    \begin{overpic}[width=0.5\textwidth]{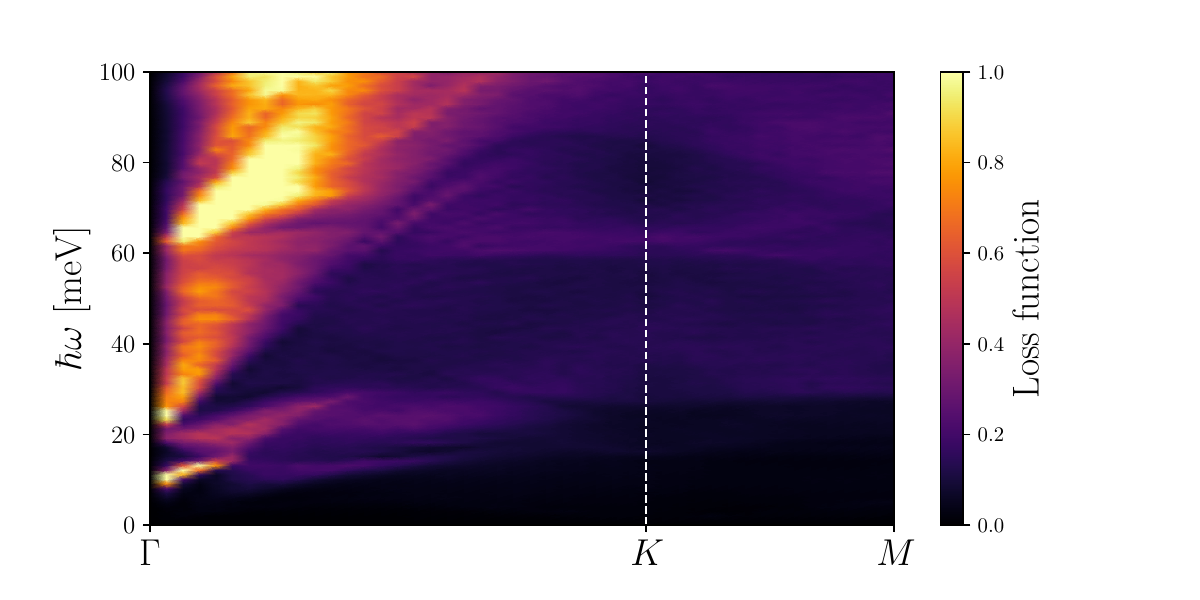}%
    \put(0,43){(a)}
    \end{overpic}
    \begin{overpic}[width=0.5\textwidth]{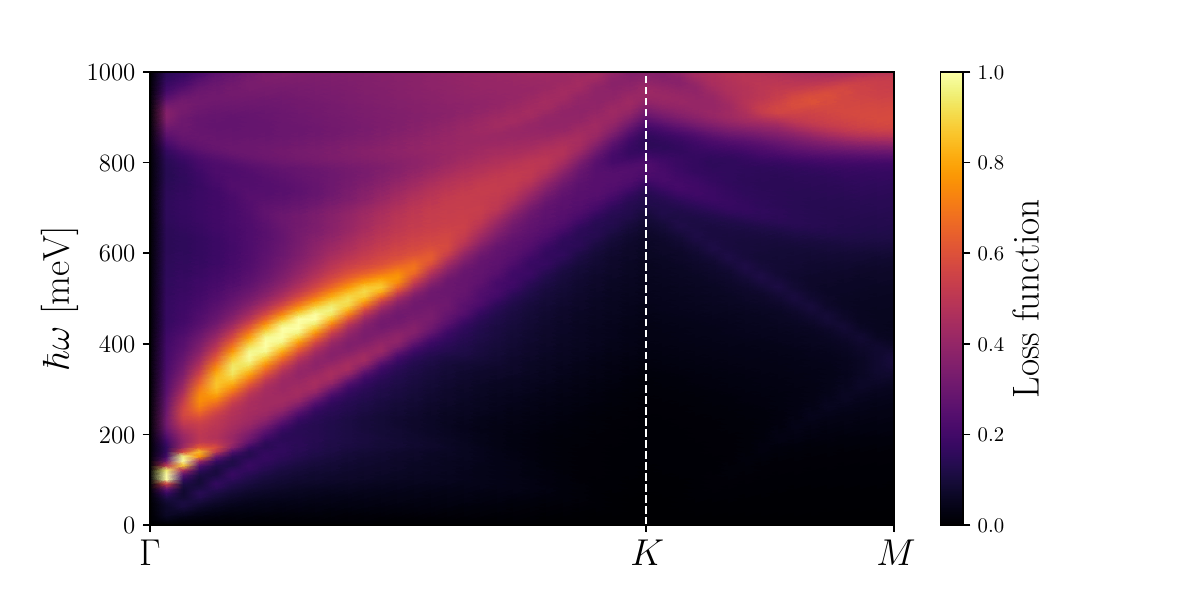}%
    \put(0,43){(b)}
    \end{overpic}
\end{tabular}
\caption{(Color online) The TBG energy loss function ${\cal L}({\bm q}, \omega)$ is plotted as a function of ${\bm q}$ and $\omega$ for $\nu=+2$. Results in this figure have been obtained by taking into account both LFEs and Hartree corrections. Panel (a) $\theta=1.05^{\degree}$. Panel (b) $\theta=5^{\degree}$. On the horizontal axis we report ${\bm q}$ along the high-symmetry path $\Gamma$-$K$-$M$ of the moir\'{e} BZ---see Fig.~\ref{fig:sketch+mBZ}(b) in the main text.\label{fig:acoustic_appearence2}}
\end{figure}

We evaluated the energy loss function ${\cal L}({\bm q}, \omega)$  with LFEs and Hartree corrections (while taking into account the layer-pseudospin degree of freedom) for $\theta = 1.05^{\degree}$ and $\theta = 5^{\degree}$, at filling factor $\nu = +2$. Results are shown in Figure~\ref{fig:acoustic_appearence2}. Similarly to Fig.~\ref{fig:acoustic_appearence} (which refers to $\nu=+1$), we clearly see an intrinsic acoustic plasmon mode for $\theta = 5^{\degree}$. Note that the peak in the energy loss function associated to the acoustic plasmon is weaker for $\nu=+2$ than $\nu=+1$.

\begin{figure*}
\begin{tabular}{ll}
    \begin{overpic}[width=0.5\textwidth]{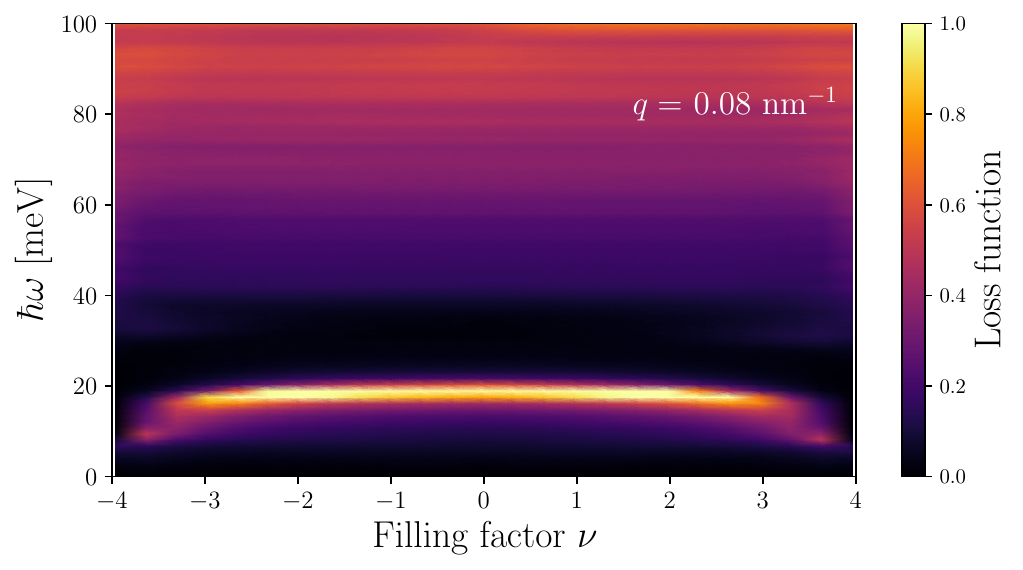}%
    \put(0,53){(a)}
    \end{overpic} & \begin{overpic}[width=0.5\textwidth]{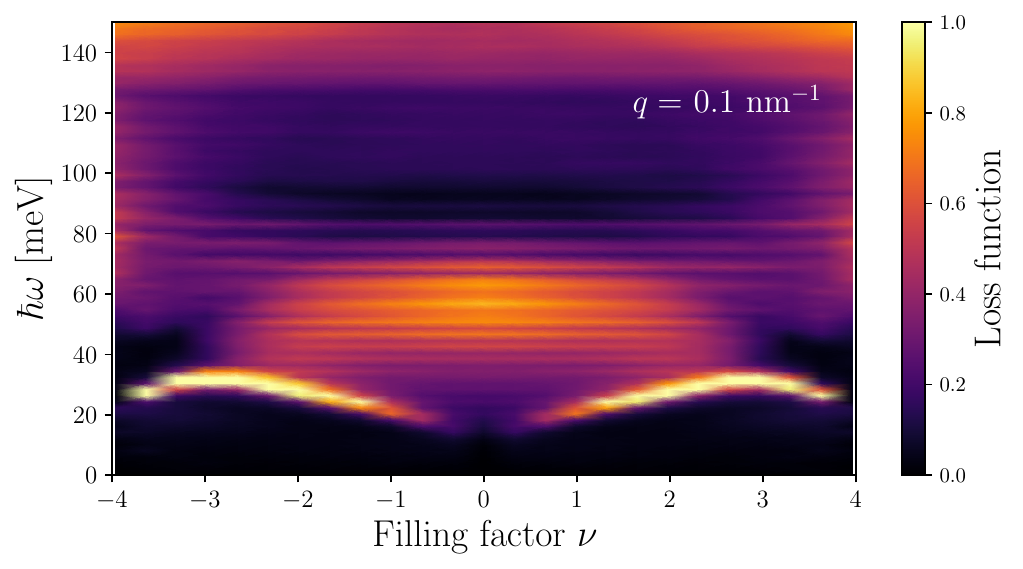}%
    \put(0,53){(b)}
    \end{overpic}\\
    \begin{overpic}[width=0.5\textwidth]{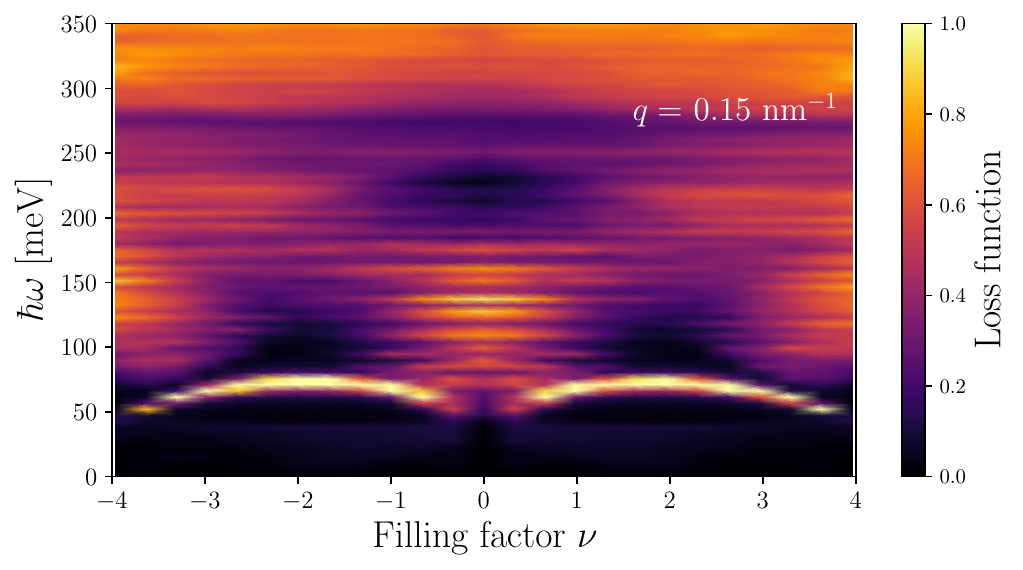}%
    \put(0,53){(c)}%
    \end{overpic} & \begin{overpic}[width=0.5\textwidth]{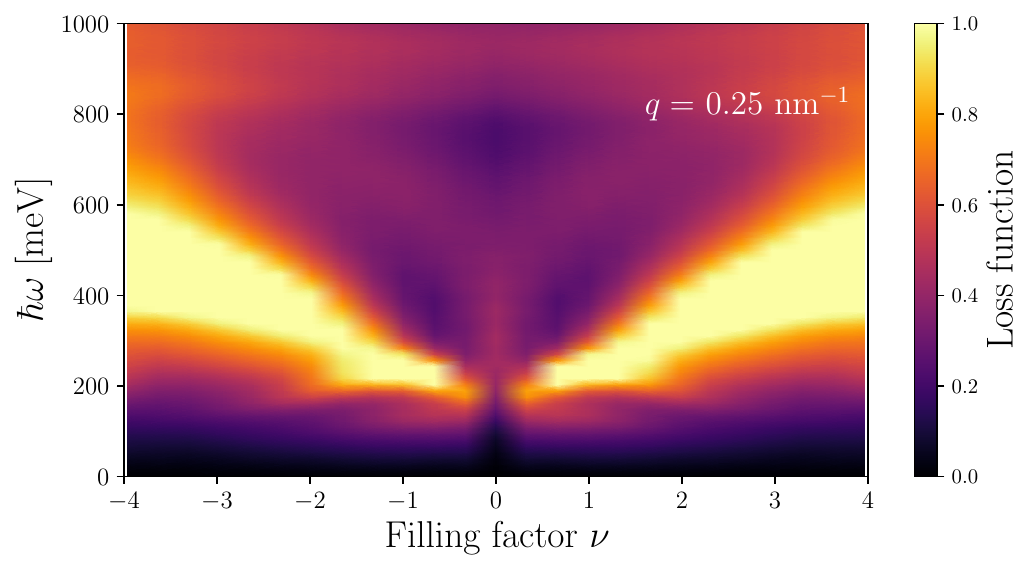}%
    \put(0,53){(d)}%
    \end{overpic}\\
\end{tabular}
\caption{(Color online) The TBG energy loss function ${\cal L}({\bm q}, \omega)$ is plotted as a function of $\omega$ and filling factor $\nu$. Each panel corresponds to a value of $\theta$ and $q$ (the latter taken along the $\Gamma$-$K$ direction of the moir\'{e} BZ). Results in this figure have been obtained by {\it neglecting} Hartree corrections. Panel (a) $\theta = 1.05^{\degree}$. Panel (b) $\theta=1.35^{\degree}$. Panel (c) $\theta = 2^{\degree}$. Panel (d) $\theta=5^{\degree}$.\label{fig:loss_function_fillings}}
\end{figure*}

\begin{figure*}
\begin{tabular}{ll}
    \begin{overpic}[width=0.5\textwidth]{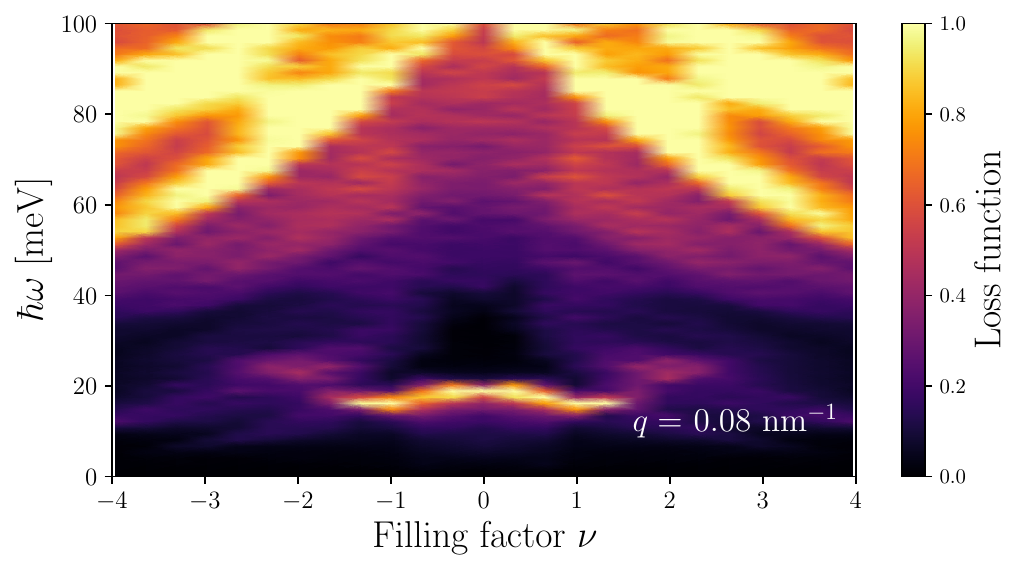}%
    \put(0,53){(a)}
    \end{overpic} & \begin{overpic}[width=0.5\textwidth]{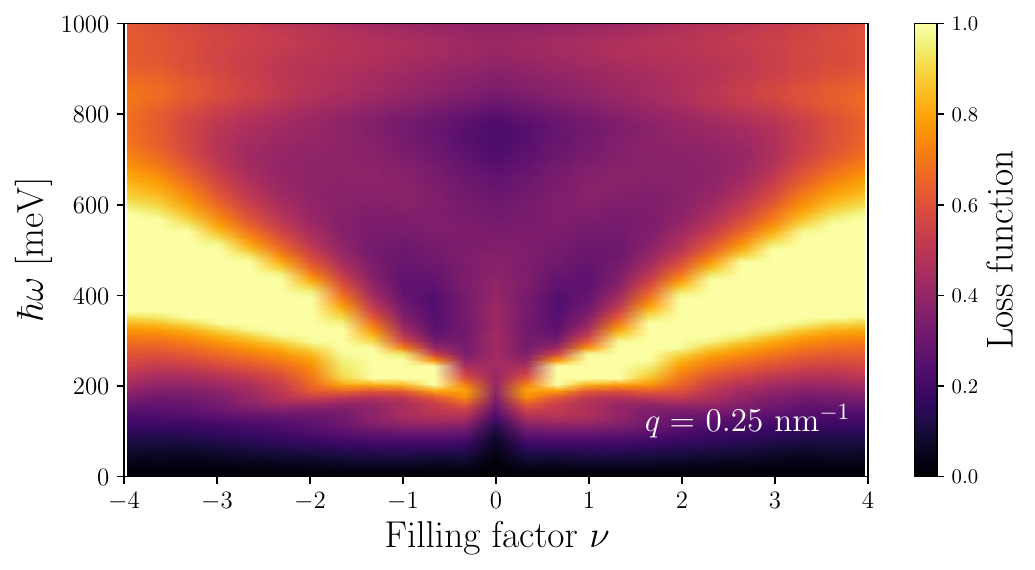}%
    \put(0,53){(b)}
    \end{overpic}
\end{tabular}
\caption{(Color online) The TBG energy loss function ${\cal L}({\bm q}, \omega)$ is plotted as a function of $\omega$ and filling factor $\nu$. Each panel corresponds to a value of $\theta$ and $q$ (the latter taken along the $\Gamma$-$K$ direction of the moir\'{e} BZ). Results in this figure have been obtained by {\it including} Hartree corrections. Panel (a) $\theta = 1.05^{\degree}$. Panel (b) $\theta=5^{\degree}$. \label{fig:loss_function_fillings2}}
\end{figure*}

Figure~\ref{fig:loss_function_fillings} shows the energy loss function for different values of $\nu$ and fixed values of the wave vector ${\bm q}$ taken along the $\Gamma$-$K$ direction in the moir\'{e} BZ. Results in Fig.~\ref{fig:loss_function_fillings} have been obtained by neglecting the Hartree contribution. In Figure~\ref{fig:loss_function_fillings2} we show similar results---for two values of $\theta$, i.e.~$\theta = 1.05^{\degree}$ and $\theta= 5^{\degree}$---but this time with the inclusion of the Hartree contribution. Note the very high level of particle-hole symmetry in the plasmonic spectrum and a significant suppression of the acoustic plasmon for $|\nu|>2$.
\section*{Section IV: Effects of a static, perpendicular electric field}

In this Section, we discuss the effect of a static, perpendicular electric field $\bm{E}_{z} = E_{0}\bm{z}$ on the plasmonic spectrum of TBG. Within the continuum model introduced in Sect.~\ref{subsect:TBG bare-band model} of the main text, such an electric field is taken into account by adding to the Hamiltonian (\ref{eqn:continuumtot}) the following contribution:
\begin{equation}\label{eq:Helectric_field}
    \hat{\cal H}_{\rm el} = U_{\rm el}\left(\hat{\Pi}^{(1)}-\hat{\Pi}^{(2)}\right)~.
\end{equation}
Here, $\hat{\Pi}^{(i)}$ is the projector on the $i$-th layer defined in Eq.~\eqref{eqn:projector_def} of the main text and
\begin{equation}\label{eq:electrical_potential_energy}
U_{\rm el} = \frac{1}{2}e E_0 d~,    
\end{equation}
where $e$ is the elementary charge and $d \approx 0.3~{\rm nm}$ the spatial separation between the two graphene layers. 

\begin{figure*}
\begin{tabular}{lll}
    \begin{overpic}[width=0.5\textwidth]{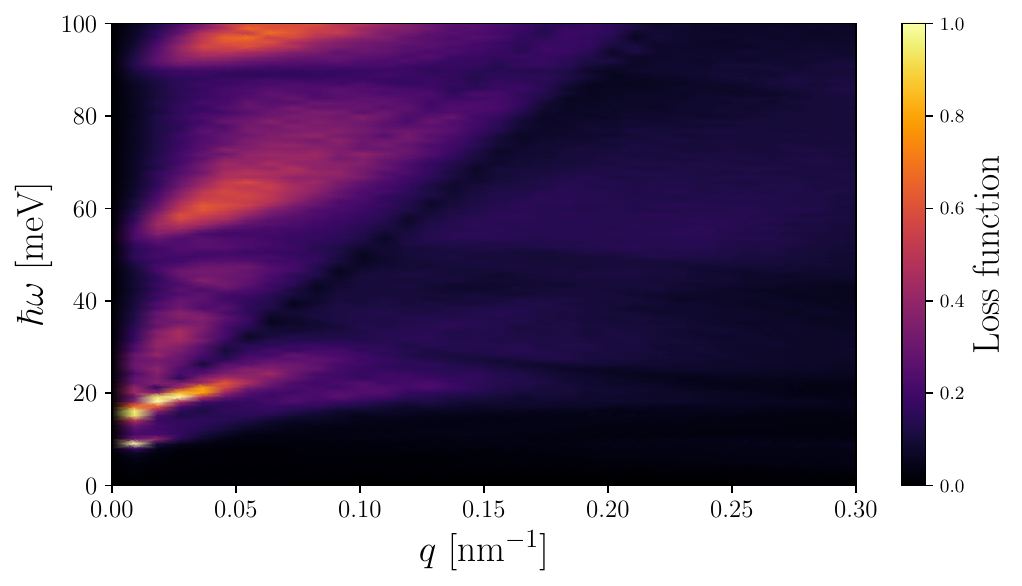}%
    \put(0,53){(a)}
    \end{overpic} & \begin{overpic}[width=0.5\textwidth]{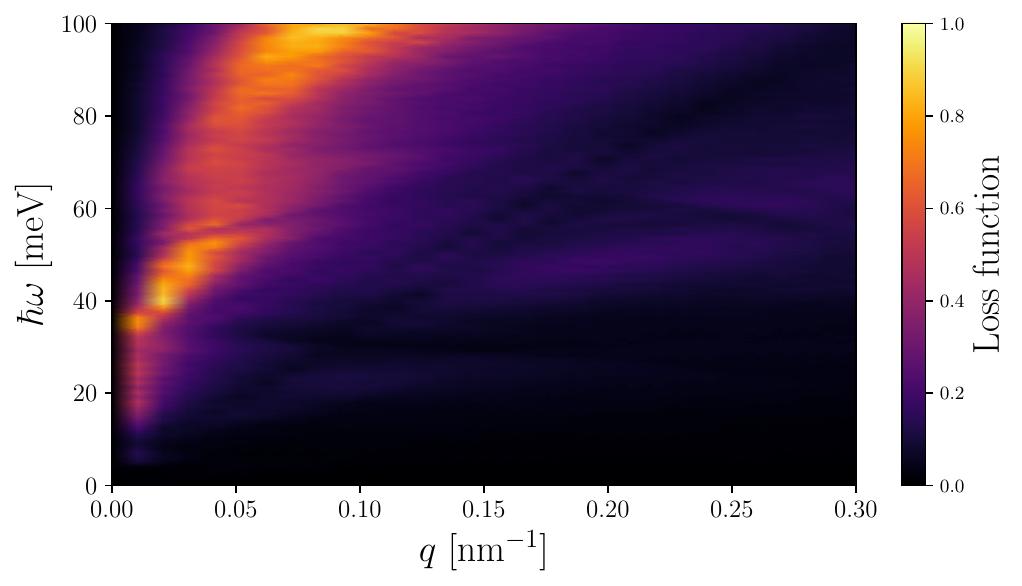}%
    \put(0,53){(b)}
    \end{overpic}\\
    \begin{overpic}[width=0.5\textwidth]{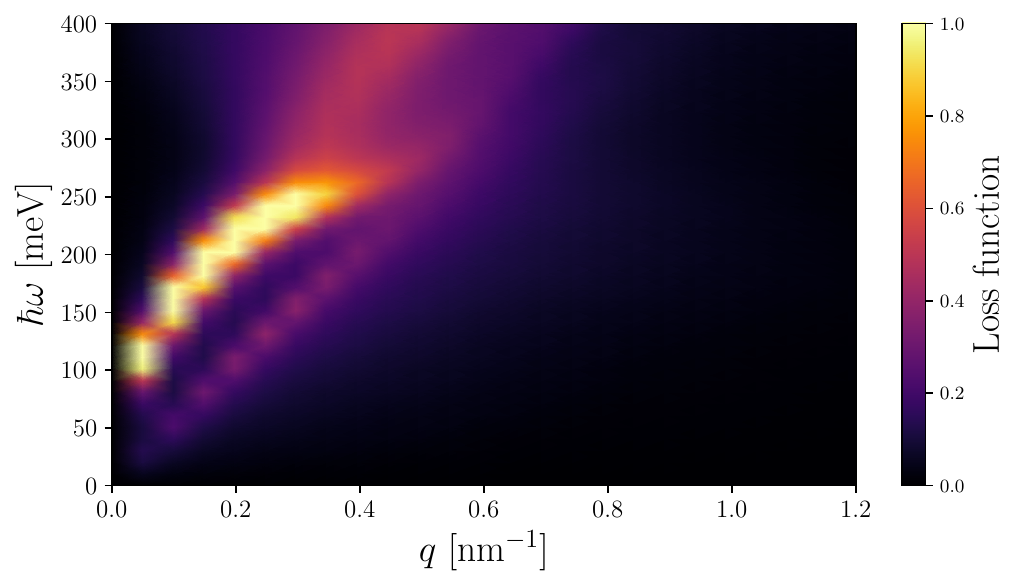}%
    \put(0,53){(c)}%
    \end{overpic} & \begin{overpic}[width=0.5\textwidth]{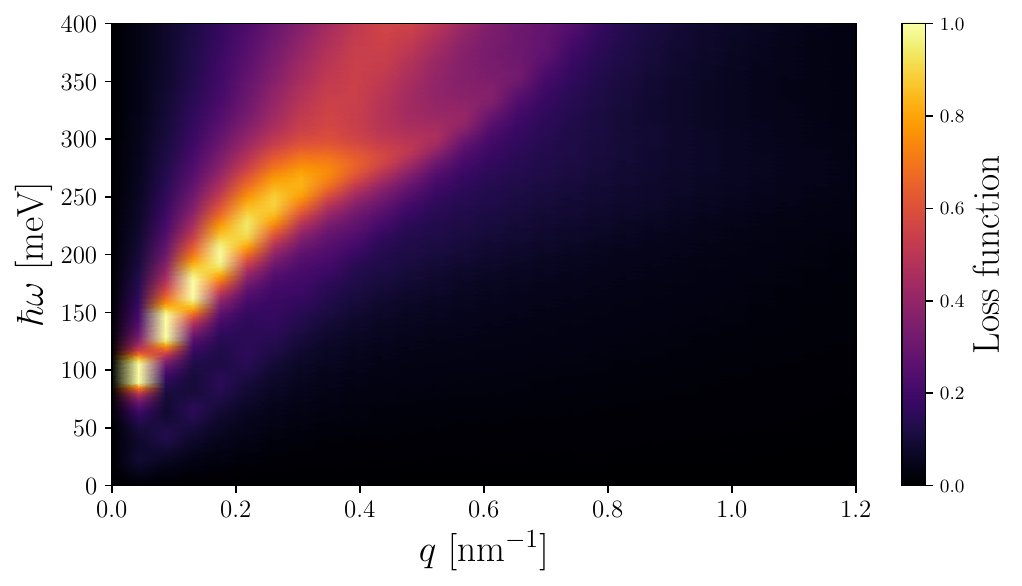}%
    \put(0,53){(d)}%
    \end{overpic}
\end{tabular}
\caption{(Color online) Dependence of the plasmonic spectrum on a static, perpendicular electric field. The TBG energy loss function ${\cal L}({\bm q},\omega)$ is plotted as a function of ${\bm q}$ and $\omega$ for two values of $U_{\rm el}$: see Eq.~(\ref{eq:electrical_potential_energy}). On the horizontal axis we report ${\bm q}$ along the high-symmetry path $\Gamma$-$K$ of the moir\'{e} BZ: see Fig.~\ref{fig:sketch+mBZ}(b) in the main text. The left (right) column refers to $E_0 = 0.5~{\rm V/nm}$ ($E_0 = 1~{\rm V/nm}$). Panels (a)-(b): $\theta=1.05^{\degree}$. At this small twist angle, the applied electric field suppresses the COM mode. Panels (c)-(d): $\theta=5^{\degree}$. At this larger value of the twist angle, instead, the main effect of the electric field is to suppress the acoustic mode.  \label{fig:loss_vs_Ez}}
\end{figure*}

We have evaluated the impact of Eq.~(\ref{eq:Helectric_field}) on the energy loss function ${\cal L}(\bm{q}, \omega)$, by including LFEs and Hartree corrections. A summary of our main findings is reported in Fig.~\ref{fig:loss_vs_Ez}, where we show results obtained for $\theta = 1.05^{\degree}$---panels (a) and (b)---and~$\theta = 5^{\degree}$---panels (c) and (d). For each of the two twist angles, we considered two values of the electric field, namely $E_0 = 0.5~{\rm V/nm}$ and $E_0 = 1~{\rm V/nm}$ which correspond to a potential energy of $U_{\rm el} \approx 84~{\rm meV}$ and $U_{\rm el} = 168~{\rm meV}$, respectively. Clearly, the impact of the perpendicular electric field is more pronounced at small angles, as it suppresses the COM plasmon already for $E_0 = 0.5~{\rm V/nm}$: see Fig.~\ref{fig:loss_vs_Ez}(a). At higher angles, instead, the plasmonic spectrum is only slightly modified, Fig.~\ref{fig:loss_vs_Ez}(c)-(d), the main effect of a large applied electric field being the suppression of the acoustic plasmon. 

These results can be qualitatively explained as following. The applied perpendicular electric field polarizes TBG, leading to charge accumulation onto one of two layers and a consequent depletion of charge in the other layer. It is precisely this imbalance that is at the origin of the suppression of the COM mode at  small twist angles, where the two layers are strongly tunnel-coupled, and of the acoustic plasmon at large twist angles, where the two layers are weakly tunnel-coupled.
\section*{Section V: Heterostrain effects on plasmons}

In this Section, we explore the effects of heterostrain on the plasmonic spectrum of TBG. We first briefly review how strain modifies the reciprocal lattice of the bilayer system. We then introduce the continuum model~\cite{SM_bi_prb_2019} that describes heterostrained TBG.

Heterostrain refers to relative strains between layers, and can be present in experimental TBG samples either because of unwanted interactions with the substrate or because intentionally applied and controlled by piezoelectrics. The properties of heterostrained TBG can be captured, in the small deformation and small rotation limit, by the following deformation matrices:
\begin{equation}\label{eqn:deformation_matrix}
    {\cal E}^{(\ell)} = \left(\begin{array}{cc}
        \epsilon_{xx}^{(\ell)} & \epsilon_{xy}^{(\ell)} - (-)^{\ell}\theta/2 \\
        \epsilon_{yx}^{(\ell)} + (-)^{\ell} \theta/2 & \epsilon_{yy}^{(\ell)}
    \end{array}\right)~.
\end{equation}
The strained geometry leads to a deformed moir\'e reciprocal lattice. This can be constructed according to the following equation,
\begin{equation}\label{eqn:reciprocal_lattice_basis_strain}
    \tilde{\bm G}_i = {\cal E}^{\rm T} \bm{g}_i~,
\end{equation}
where $\bm{g}_1 = \frac{4\pi}{\sqrt{3} a}(\frac{\sqrt{3}}{2}, -\frac{1}{2})$ and $\bm{g}_2 = \frac{4\pi}{\sqrt{3} a}(0, 1)$ are the unstrained and untwisted reciprocal lattice vectors and ${\cal E} \equiv {\cal E}^{(2)} - {\cal E}^{(1)}$ is the relative deformation matrix. The quantity $\tilde{\bm G}_i$ defines the strained moir\'e reciprocal lattice counterpart of the unstrained reciprocal lattice obtained by the vectors defined in the main text in Eq.~\eqref{eqn:reciprocal_lattice_basis}. 

In what follows, we further assume ${\cal E}^{(2)} = -{\cal E}^{(1)} = \frac{1}{2}{\cal E}$, as in Ref.~\cite{SM_bi_prb_2019} and limit our investigation to uniaxial heterostrain. This type of heterostrain involves the application of stress predominantly along one direction of the bilayer system while leaving the perpendicular direction unstressed. With all these restrictions, the strain part of the relative deformation matrix ${\cal E}$ can be expressed with only three parameters: strain magnitude $\epsilon$, strain direction $\phi$, and Poisson ration $\nu_{\rm p}$, which takes the value $\nu_{\rm p}\approx0.16$ in graphene. We find
\begin{equation}
    {\cal E} = {\cal R}^{-1}(\phi)\left(\begin{array}{cc}
        -\epsilon & 0 \\
        0 & \nu_{\rm p}\epsilon
    \end{array}\right){\cal R}(\phi) + \left(\begin{array}{cc}
        0 & -\theta \\
        \theta & 0
    \end{array}\right)~,
\end{equation}
where the rotation matrix ${\cal R}(\phi)$ is given by
\begin{equation}
    {\cal R}(\phi) = \left(\begin{array}{cc}
        \cos\phi& -\sin\phi \\
        \sin\phi & \cos\phi
    \end{array}\right)~.
\end{equation}

\begin{figure*}
\begin{tabular}{lll}
    \begin{overpic}[width=0.5\textwidth]{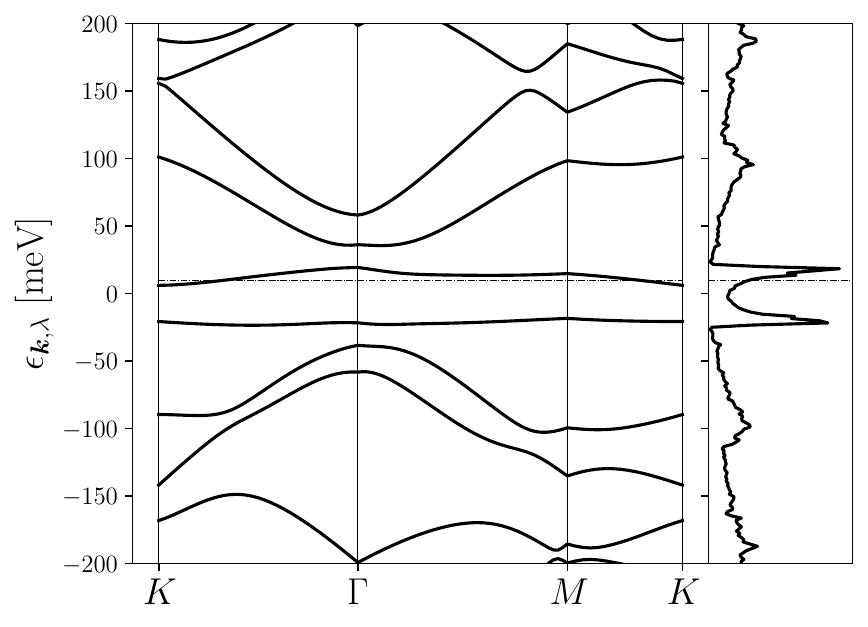}%
    \put(0,70){(a)}
    \end{overpic} & \begin{overpic}[width=0.5\textwidth]{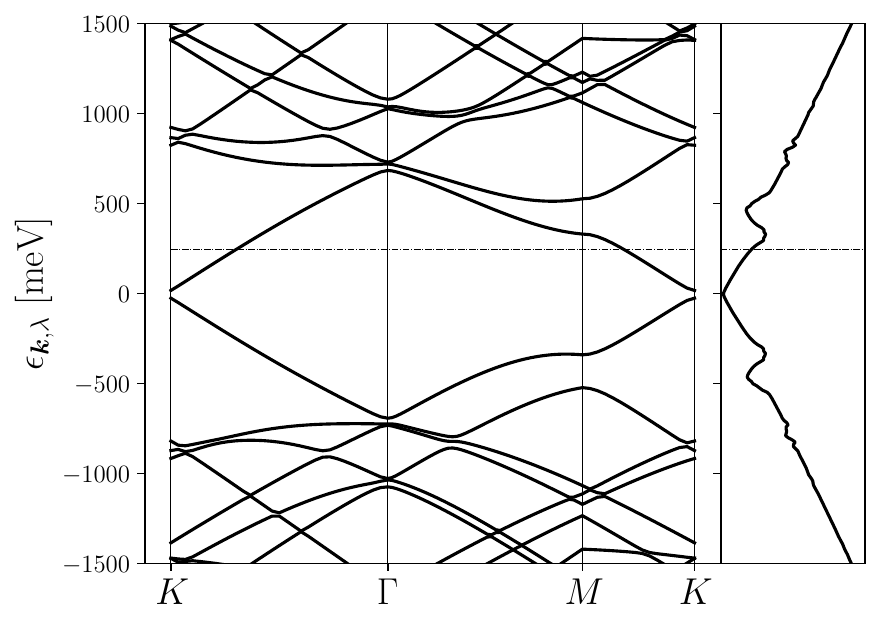}%
    \put(0,70){(b)}
    \end{overpic}
\end{tabular}
\caption{(Color online) The energy bands of TBG and their corresponding density of states calculated by taking into account uniaxial heterostrain. On the horizontal axis we report the momentum ${\bm k}$ along the high-symmetry path $K$-$\Gamma$-$M$-$K$ of the strained moir\'{e} BZ. In order to obtain these results we fixed $\varepsilon =  0.6\%$ and $\phi = 30^{\degree}$. In panel (a) the twist angle is $\theta = 1.05^{\degree}$, while in panel (b) $\theta = 5^{\degree}$. In both panels the dashed-dot line shows the chemical potential $\mu$ for filling factor $\nu = +1$ and temperature $T = 5$ K.\label{fig:bands_vs_strain}}
\end{figure*}

\begin{figure*}
\begin{tabular}{lll}
    \begin{overpic}[width=0.5\textwidth]{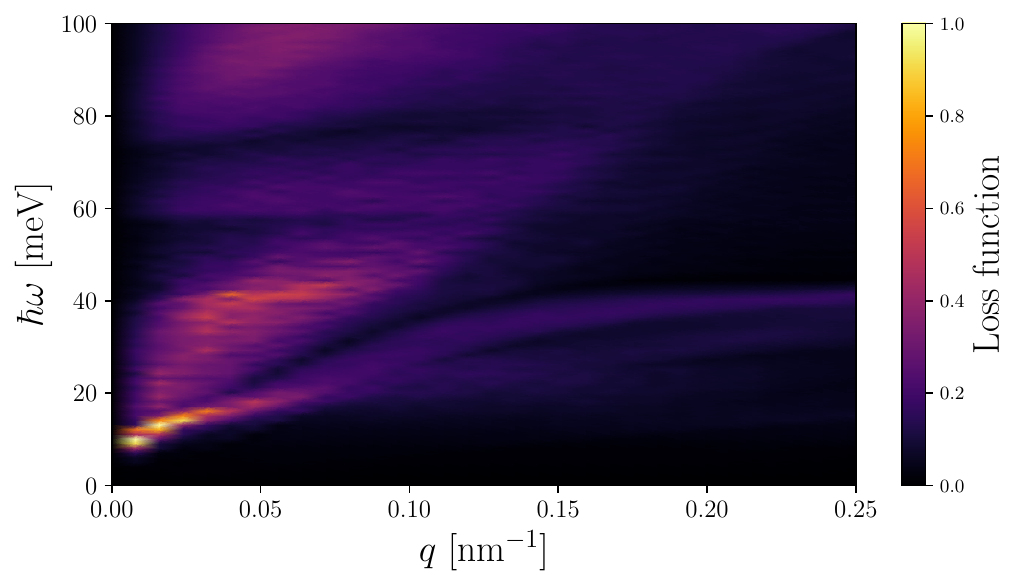}%
    \put(0,53){(a)}
    \end{overpic} & \begin{overpic}[width=0.5\textwidth]{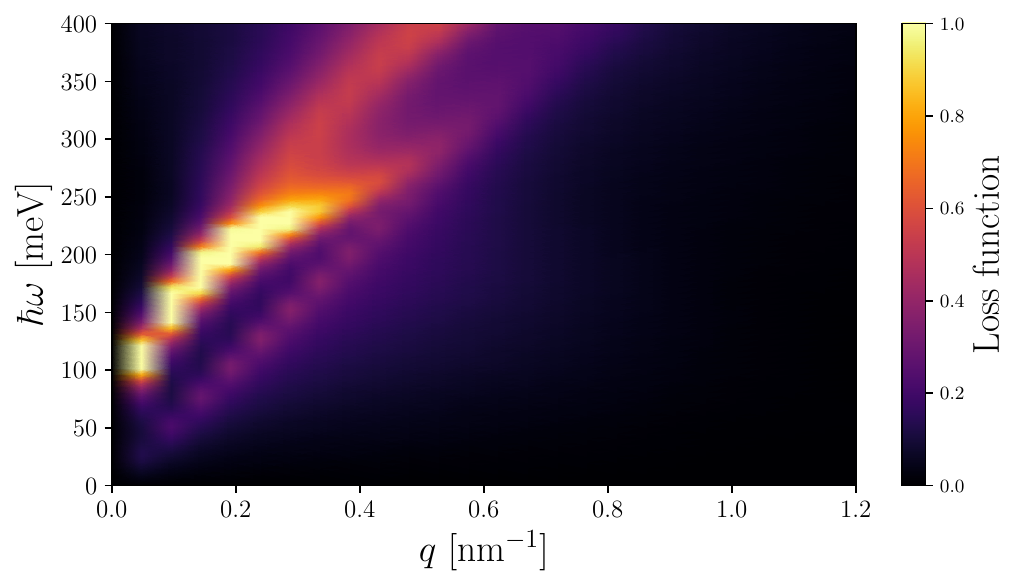}%
    \put(0,53){(b)}
    \end{overpic}
\end{tabular}
\caption{(Color online) The impact of uniaxial heterostrain on the TBG energy loss function ${\cal L}({\bm q},\omega)$, which is plotted, as usual, as a function of ${\bm q}$ (along the high-symmetry path $\Gamma$-$K$ of the strained mBZ) and $\omega$. Results in this figure have been obtained by choosing $\epsilon = 0.6 \%$ and $\phi=30^{\degree}$. Panel (a): $\theta=1.05^{\degree}$. We clearly see that heterostrain largely suppresses the COM plasmon while pushing the
inter-band transitions toward lower energies. Panel (b): $\theta=5^{\degree}$. In this case heterostrain does not affect the plasmonic spectrum as a direct comparison with e.g. Fig.~\ref{fig:acoustic_appearence_appendix} shows. In both panels the filling factor is fixed at $\nu = +1$ and temperature is $T = 5$ K.\label{fig:loss_vs_strain}}
\end{figure*}

\begin{figure*}
\begin{tabular}{lll}
    \begin{overpic}[width=0.5\textwidth]{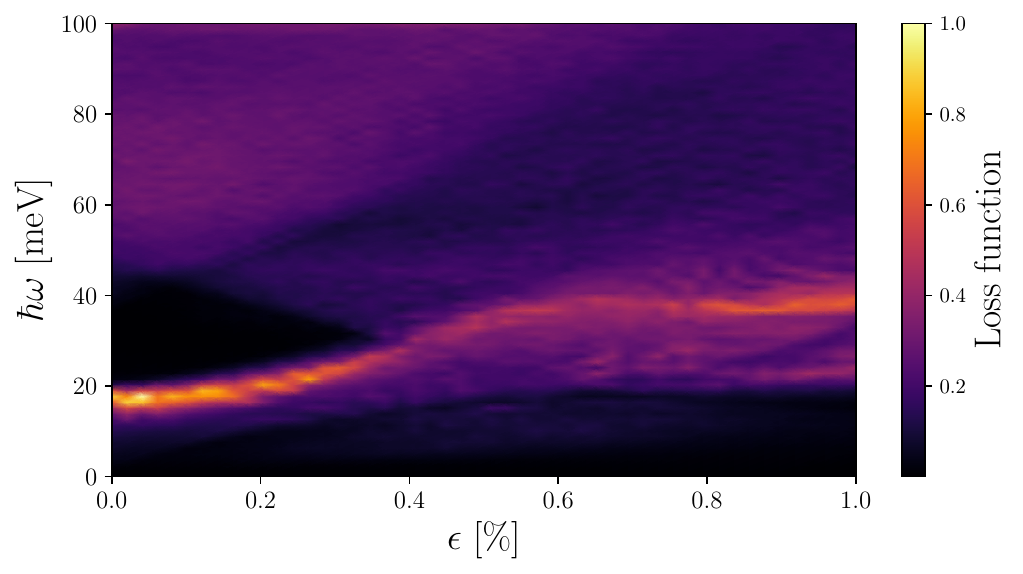}%
    \put(0,53){(a)}
    \end{overpic} & \begin{overpic}[width=0.5\textwidth]{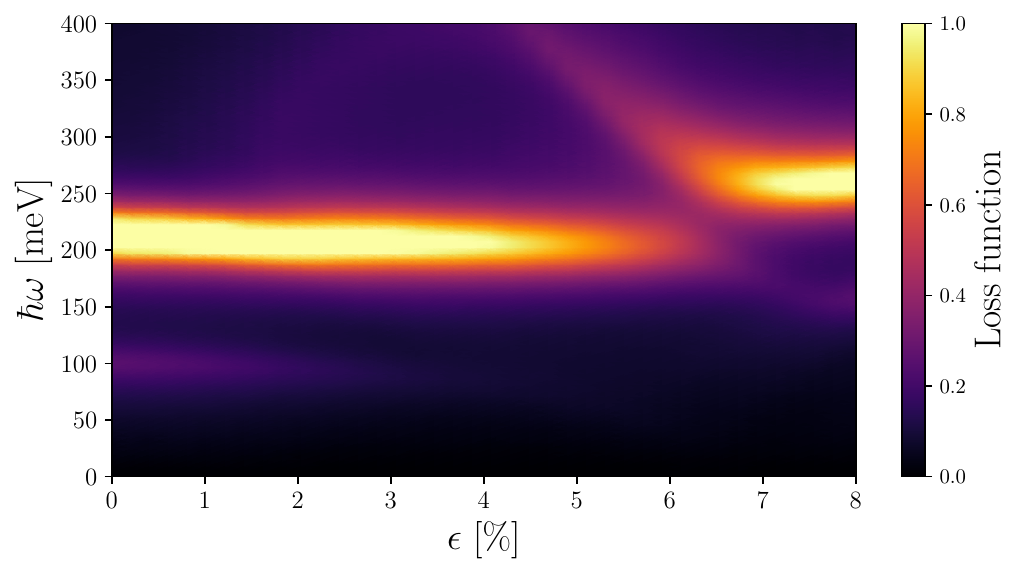}%
    \put(0,53){(b)}
    \end{overpic}
\end{tabular}
\caption{(Color online) The TBG energy loss function ${\cal L}({\bm q},\omega)$ is plotted as a function of $\hbar\omega$ and strain magnitude $\epsilon$. Results in this figure have been obtained by setting $q = \xi |\bm{K}_{1,2}|$ (see Sect.~V), $\phi = 0^{\degree}$, $\nu = +1$, and $T = 5~{\rm K}$. Panel (a): $\theta=1.05^{\degree}$. We clearly see that for $\epsilon\gtrsim0.4\%$ the COM mode is suppressed and merges with inter-band plasmons. Panel (b): $\theta=5^{\degree}$. We clearly see that the acoustic plasmon, which in this figure is the feeble feature at $\hbar\omega\approx 100~{\rm meV}$, is visible for $\epsilon \lesssim 2.5\%-3.0\%$. \label{fig:loss_vs_more_strain}}
\end{figure*}

We now move on to describe the continuum model Hamiltonian for the uniaxial heterostrained TBG. The structure of the Hamiltonian operator is the same as we described in Sect.~\ref{sec:TBG_model} of the main text, with some modifications. Within the two-center approximation, the effect of strain on the intra-layer Hamiltonian can be described by an effective vector potential (gauge field)~\cite{SM_bi_prb_2019, SM_tam_prb_2017}:
\begin{equation}\label{eqn:continuumintra_strain}
    \hat{{\cal H}}^{(\ell)}_{\xi} = v_{\rm D}\left[\left(\mathbb{I} + {\cal E}^{\rm T}\right)(\hat{\bm{p}} - \hbar \tilde{\bm{K}}_{\xi,\ell} + \xi \bm{A}_\ell)\right]\cdot(\xi\sigma_x,-\sigma_y)~.
\end{equation}
Here $\xi=\pm$ is the valley index, $\tilde{\bm{K}}_{\xi,\ell}$ is the $K$ point of the strained mBZ, and $\bm{A}_\ell$ is the effective vector potential defined as:
\begin{equation}
    \bm{A}_{2} = -\bm{A}_1 = \frac{\sqrt{3}}{4a} \beta\epsilon (1 + \nu_{\rm p})\left(\cos(2\phi), \sin(2\phi)\right)~,
\end{equation}
with $\beta\approx 3.14$ in graphene. Concerning inter-layer tunneling, it should be expressed in terms of the strained reciprocal lattice basis vectors introduced in Eq.~\eqref{eqn:reciprocal_lattice_basis_strain}. With respect to the main text expression,~cf. Eq.~\eqref{eqn:interlayeroperator}, we have a slightly different formula that takes into account the different reciprocal lattice basis:
\begin{equation}
    \hat{U} = \left(\begin{array}{cc}
       u_0  & u_1 \\
       u_1  & u_0
    \end{array}\right) + e^{i\xi\tilde{\bm{G}_1}\cdot\hat{\bm{r}}} \left(\begin{array}{cc}
       u_0  & u_1 e^{-i\frac{2\pi}{3}}\\
       u_1e^{i\frac{2\pi}{3}}  & u_0
    \end{array}\right) + e^{i\xi\left(\tilde{\bm G}_1+\tilde{\bm G}_2\right)\cdot\hat{\bm{r}}} \left(\begin{array}{cc}
       u_0  & u_1 e^{i\frac{2\pi}{3}}\\
       u_1e^{-i\frac{2\pi}{3}}  & u_0
    \end{array}\right)~.
\end{equation}
Moir\'e minibands as modified by uniaxial heterostrain are displayed in Fig.~\ref{fig:bands_vs_strain}. Results in this figure refer to $\epsilon = 0.6\%$ and strain direction $\phi = 30^{\degree}$. The impact of uniaxial heterostrain on the energy loss function is illustrated in Figs.~\ref{fig:loss_vs_strain} and~\ref{fig:loss_vs_more_strain}. In Fig.~\ref{fig:loss_vs_strain} the energy loss function is displayed for $\epsilon = 0.6\%$ and $\phi=30^{\degree}$. Instead, in Fig.~\ref{fig:loss_vs_more_strain}, we show results for the energy loss function at fixed momentum $q = \xi |\bm{K}_{1,2}|$, where $|\bm{K}_{1}|= |{\bm K}_{2}|$ is the modulus of the vector linking $\Gamma$ to $K$ in the moir\'e BZ (which depends on the strain magnitude $\epsilon$). In Fig.~\ref{fig:loss_vs_more_strain} we used $\xi = 4/25$, and varied $\epsilon$ at fixed $\phi=0^{\degree}$. In order to obtain these plots, Hartree contributions have been neglected, while we have retained LFEs. These results suggest that, at small twist angles, heterostrain largely suppresses the COM plasmon, which becomes extremely feeble and flat, while it pushes inter-band excitations toward lower energies: see Figs.~\ref{fig:loss_vs_strain}(a) and~\ref{fig:loss_vs_more_strain}(a). At larger angles, instead, the effect of strain is less severe, suggesting resilience of the acoustic plasmon for $\epsilon \lesssim 2.5\%$, as show in Figs.~\ref{fig:loss_vs_strain}(b) and~\ref{fig:loss_vs_more_strain}(b).

%

%
\end{document}